\documentstyle[12pt,epsf]{article}

\renewcommand{\textfraction}{0}

\def\la{\langle}
\def\ra{\rangle}
\def\eps{\varepsilon}

\def\l{\left}
\def\r{\right}

\makeatletter
\def\textfraction{0.2}
\@textmin = \textfraction\textheight
\def\fps@figure{tbp} % Changed RBSz default placement for figures
\def\fps@table{tbp} % Changed RBSz default placement for tables
\makeatother

\begin{document}

\begin{titlepage}

\begin{flushright}
CERN-TH/97-91\\
CPT-97/P3480\\
hep-ph/9712417
\end{flushright}

\vspace{1.0cm}
\boldmath
\begin{center}
\Large\bf
Dispersive Bounds on the Shape of\\
$\bar B\to D^{(*)}\ell\,\bar\nu$ Form Factors
\end{center}
\unboldmath

\vspace{0.3cm}
\begin{center}
Irinel Caprini\\[0.1cm]
{\sl Institute of Atomic Physics, Bucharest, POB MG-6, Romania}\\
[1.0cm]
Laurent Lellouch$^{*)}$ and Matthias Neubert\\[0.1cm]
{\sl Theory Division, CERN, CH-1211 Geneva 23, Switzerland}
\end{center}

\vspace{0.3cm}
\begin{abstract}
\vspace{0.2cm}\noindent
Dispersive constraints on the shape of the form factors which describe
the exclusive decays $\bar B\to D^{(*)}\ell\,\bar\nu$ are derived by
fully exploiting spin symmetry in the ground-state doublet of
heavy--light mesons. The analysis includes all twenty $\bar B^{(*)}\to
D^{(*)}$ semileptonic form factors. Heavy-quark symmetry, with both
short-distance and $1/m_Q$ corrections included, is used to provide
relations between the form factors near zero recoil. Simple
one-parameter functions are derived, which describe the form factors in
the semileptonic region with an accuracy of better than 2\%. The
implications of our results for the determination of $|V_{cb}|$ are
discussed.
\end{abstract}

\vspace{0.5cm}
\centerline{(Submitted to Nuclear Physics B)}

\vspace{2.0cm}\noindent
CERN-TH/97-91\\
December 1997

\vfill\noindent
$^{*)}$ {\small On leave from: Centre de Physique Th\'eorique, Case
907, CNRS Luminy, F-13288 Marseille Cedex 9, France (UPR7061).}

\end{titlepage}

\section{Introduction}

The differential rates for the exclusive semileptonic decays
$\bar B\to D^{(*)}\ell\,\bar\nu$ are given by \cite{Vcb}
\begin{eqnarray}
   \frac{\mbox{d}\Gamma(\bar B\to D^*\ell\,\bar\nu)}{\mbox{d}w}
   &=& \frac{G_F^2|V_{cb}|^2}{48\pi^3}\,(m_B-m_{D^*})^2\,m_{D^*}^3
    \sqrt{w^2-1}\,(w+1)^2 \nonumber\\
   &&\times \bigg[ 1 + \frac{4w}{w+1}\,
    \frac{m_B^2-2w\,m_B m_{D^*} + m_{D^*}^2}{(m_B - m_{D^*})^2}
    \bigg]\,|{\cal F}(w)|^2 \,, \nonumber\\[0.4cm]
   \frac{\mbox{d}\Gamma(\bar B\to D\,\ell\,\bar\nu)}{\mbox{d}w}
   &=& \frac{G_F^2|V_{cb}|^2}{48\pi^3}\,(m_B+m_D)^2\,m_D^3
    (w^2-1)^{3/2} |V_1(w)|^2 \,,
\label{BDrate}
\end{eqnarray}
where ${\cal F}(w)$ and $V_1(w)$ are hadronic form factors, and
$w=v_B\cdot v_{D^{(*)}}$ is the product of the velocities of the
initial and final mesons. In the heavy-quark limit, ${\cal F}(w)$ and
$V_1(w)$ coincide with the Isgur--Wise function $\xi(w)$, which
describes the long-distance physics associated with the light degrees
of freedom in the heavy mesons \cite{Isgu}. This function is normalized
to unity at zero recoil, corresponding to $w=1$. Corrections to this
limit can be calculated using the heavy-quark effective theory
\cite{Review}, and are suppressed by powers of $\alpha_s(m_Q)$ or
$\Lambda_{\rm QCD}/m_Q$, where we use $m_Q$ generically for $m_b$ or
$m_c$. Detailed calculations of these corrections lead to ${\cal
F}(1)=0.91\pm 0.03$ \cite{Vcb}, \cite{Luke}--\cite{Czar} and
$V_1(1)=0.98\pm 0.07$ \cite{LNN}, so that an accurate determination of
$|V_{cb}|$ can be obtained by extrapolating the differential decay
rates to $w=1$. To reduce the uncertainties associated with this
extrapolation, constraints on the shape of the form factors are highly
desirable. A suitable framework to derive such constraints is provided
by a dispersion technique proposed some time ago
\cite{Meim}--\cite{Bour}, and has been applied to heavy-meson form
factors more recently \cite{Taro1}--\cite{Boyd3}. This method is based
on first principles: the analyticity properties of two-point functions
of local currents and the positivity of the corresponding spectral
functions. Analyticity relates integrals of the hadronic spectral
functions to the behavior of QCD two-point functions in the deep
Euclidean region via dispersion relations. Positivity guarantees that
the contributions of the states of interest to the spectral functions
are bounded above. The constraints on the relevant form factors, given
their analyticity properties, then follow from these bounds.

Following Refs.~\cite{CaMa,Boyd3}, we generalize this method to fully
exploit spin symmetry in the doublets of the ground-state $B^{(*)}$ and
$D^{(*)}$ mesons. By taking into account all contributions related by
heavy-quark symmetry, this generalization increases the constraining
power of the method considerably. We investigate all spin--parity
channels ($J^P=0^+$, $0^-$, $1^-$ and $1^+$) relevant for $\bar
B^{(*)}\to D^{(*)}$ transitions. This gives us four inequalities and
thus four allowed domains for the parameters which describe the form
factors of interest. The optimal constraints are obtained by taking the
intersection of these domains. Although these different channels were
also investigated recently by the authors of Ref.~\cite{Boyd3}, we
bring a number of improvements to their analysis and exploit the
dispersive techniques with a somewhat different emphasis. First, we
include the contributions of the $B_c$ poles to the relevant
polarization functions. Since these contributions are positive, they
reduce the domain available to the contributions of the $B^{(*)}
D^{(*)}$ states and therefore increase the constraints on the
corresponding form factors. As a second improvement, we include the
short-distance and $1/m_Q$ corrections to the heavy-quark limit in the
relations between form factors, thereby eliminating the leading
uncertainties in these relations. We further discuss the role of
remaining uncertainties and show how their inclusion is necessary to
avoid overinterpreting the results of the dispersive method. In
addition, instead of requiring rather complicated kinematical functions
and the knowledge of the singularity structure of individual form
factors, we introduce single-parameter descriptions that are simple,
low-order power-series expansions in kinematical variables, and
accurate to 2\% over the full semileptonic domain. Thus, these
parametrizations will be very useful for reducing the uncertainty in
the extraction of $|V_{cb}|$ from experimental data.

The paper is organized as follows. In the next section, we present the
derivation of the unitarity inequalities, which are the basis of the
dispersive bounds on form factors discussed in Section~\ref{sec:3}. In
Sections~\ref{sec:4} and \ref{sec:5}, we derive model-independent
bounds on the heavy-meson form factors using expansions in different
kinematical variables. We present a new method for including
corrections to the heavy-quark limit, which fully takes into account
the uncertainties in their calculation. We find that these
uncertainties have a significant effect on the bounds and, ultimately,
limit the accuracy of the dispersive approach. We also derive simple,
one-parameter descriptions of the form factors in the semileptonic
region. The phenomenological consequences of our results are discussed
in Section~\ref{sec:more}, where the reader not interested in the
technical details of our work can find a summary of our
parametrizations for the $\bar B\to D^{(*)}\ell\,\bar\nu$ form factors.
Our conclusions are given in Section~\ref{sec:ccl}. The paper also
comprises an Appendix, where we give the definitions of form factors
and details of our calculations.

\section{Dispersion relations and unitarity inequalities}
\label{sec:2}

We start with the vacuum polarization tensor
\begin{eqnarray}
   \Pi^{\mu\nu}(q) &=& i\int\!\mbox{d}^4x\,e^{iq\cdot x}\,
    \la\,0\,|\,T\{V^\mu(x),V^{\dagger\nu}(0)\}|\,0\,\ra
    \nonumber\\
   &=& (q^\mu q^\nu - g^{\mu \nu} q^2)\,\Pi_{1-}(q^2)
    + q^\mu q^\nu\,\Pi_{0^+}(q^2) \,,
\label{pimunu}
\end{eqnarray}
and a similar expression for the correlator of two axial currents with
invariant functions $\Pi_{1^+}$ and $\Pi_{0^-}$, where the subscripts
indicate the spin--parity quantum numbers of the intermediate states
contributing in the various channels. The invariant functions satisfy
subtracted dispersion relations, which we write as ($Q^2=-q^2$)
\begin{eqnarray}
   \chi_{0^\pm}(Q^2) &=& \left( -\frac{\partial}{\partial Q^2} \right)
    \left[ -Q^2\,\Pi_{0^\pm}(Q^2) \right]
    = \frac{1}{\pi} \int\limits_0^\infty\!\mbox{d}t\,
    \frac{t\,\mbox{Im}\Pi_{0^\pm}(t)}{\big(t+Q^2\big)^2} \,,
    \nonumber\\
   \chi_{1^\pm}(Q^2) &=& \frac 12 \left( -\frac{\partial}{\partial Q^2}
    \right)^2 \left[-Q^2\,\Pi_{1^\pm}(Q^2) \right]
    = \frac{1}{\pi} \int\limits_0^\infty\!\mbox{d}t\,
    \frac{t\,\mbox{Im}\Pi_{1^\pm}(t)}{\big(t+Q^2\big)^3} \,.
\label{chidr}
\end{eqnarray}
In the Euclidean region, where $Q^2>0$, these functions can be
calculated in QCD using the operator product expansion. On the other
hand, the spectral functions $\mbox{Im}\Pi_{0^\pm}(t)$ and
$\mbox{Im}\Pi_{1^\pm}(t)$ are given by unitarity relations. In the case
of the vector correlator, we have
\begin{eqnarray}
   &&(q^\mu q^\nu - g^{\mu \nu} q^2)\,
    \mbox{Im}\Pi_{1^-}(q^2+i\epsilon) + q^\mu q^\nu\,
    \mbox{Im}\Pi_{0^+}(q^2+i\epsilon) \nonumber\\
   &&= \frac 12 \sum\!\!\!\!\!\!\!\int\limits_\Gamma\,\,
    \mbox{d}\rho_\Gamma\,(2\pi)^4\,\delta^{(4)}(p_\Gamma-q)\,
    \la\,0\,|V^\mu(0)|\Gamma\ra\,\la\Gamma|V^{\dagger\nu}(0)
    |\,0\,\ra \,,
\label{unitar}
\end{eqnarray}
and a similar relation holds for the axial correlator. The spectral
functions, which are sums of positive terms, are bounded below by the
contributions of the two-particle states $|B D\ra$, $|B D^*\ra$, $|B^*
D\ra$, and $|B^* D^*\ra$, which are related through crossing symmetry
to the matrix elements relevant for semileptonic decays. Combining this
lower bound with the QCD result for the correlators, we derive
constraints on the form factors in the semileptonic region. These
constraints are further improved by using heavy-quark symmetry to
relate the matrix elements associated with these different
contributions. Note that, when extending the number of hadron states in
the unitarity sum, there is, in principle, a danger of double counting
if particles which decay via strong interactions are included. In our
case, $D^*$ accounts partially for the two-particle state $D\pi$, which
therefore must not be included separately. ($B^*$ is stable with
respect to strong decays, since $m_{B^*}<m_B+m_\pi$.)

To further increase the constraining power of the dispersive method, we
take into account the single-particle contributions of the first two
vector $B_c^*$ and pseudoscalar $B_c$ mesons.\footnote{We do not
include more massive resonances, because they lie above the $B D$
threshold and may lead to double counting.}
These contributions are determined in terms of the masses and leptonic
decay constants of these states, whose values can be calculated rather
accurately using quark models. In Table~\ref{tab:masses}, we collect
the masses and decay constants of all $B_c$ mesons below the threshold
for $B^* D^*$ production, as predicted by the model of
Ref.~\cite{Eich}. Note that the scalar and axial states, as well as
some of the vector and pseudoscalar states, have non-vanishing orbital
angular momentum and therefore vanishing decay constants. Although this
conclusion is strictly only valid in the context of quark models, we
expect orbitally excited states to have much smaller decay constants
than S-wave states, and take the conservative approach of neglecting
their contributions to the various spectral functions.

\begin{table}
\centerline{\parbox{14cm}{\caption{\label{tab:masses}\small\sl
Masses (in GeV) and decay constants (in MeV) of the lowest-lying $B_c$
states \protect\cite{Eich}}}}
\begin{center}
\begin{tabular}{c|c|cc|cc|c}
\hline\hline
$J^P$ & $0^+$ & \multicolumn{2}{c|}{$1^-$} &
 \multicolumn{2}{c|}{$0^-$} & $1^+$ \\
\hline
$n$ & $M_n$ & $M_n$ & $f_n$ & $M_n$ & $f_n$ & $M_n$ \\
\hline
1 & 6.700 & 6.337 & 497 & 6.264 & 500 & 6.730 \\
2 & 7.108 & 6.899 & 369 & 6.856 & 370 & 6.736 \\
3 &       & 7.012 &     & 7.244 &     & 7.135 \\
4 &       & 7.280 & 327 &       &     & 7.142 \\
\hline\hline
\end{tabular}
\end{center}
\end{table}

Inserting these one- and two-particle states into the unitarity sums,
and combining the resulting bounds with the dispersion relations of
(\ref{chidr}), we find
\begin{eqnarray}
   \sum_{j=1,2,3}&&
    \int\limits_{-\infty}^{-1}\!\mbox{d}w\,(w^2-1)^{1/2}
    \left( \frac{w+1}{2} \right)^2\,(1+\delta_{j2})\,
    \frac{(\beta_j^2-1)|S_j(w)|^2}
     {\Big(\beta_j^2-\frac{w+1}{2}\Big)^4}
    < \frac{64\pi^2}{n_f}\,\chi_{0^+}(0) \,, \nonumber\\[0.6cm]
   \sum_{j=1,2,3}&&
    \int\limits_{-\infty}^{-1}\!\mbox{d}w\,(w^2-1)^{3/2}\,
    (1+\delta_{j2})\,
    \frac{\beta_j^2|V_j(w)|^2}{\Big(\beta_j^2-\frac{w+1}{2}\Big)^5}
    \nonumber\\
   \mbox{}+ \sum_{j=4,5,6,7}&&
    \int\limits_{-\infty}^{-1}\!\mbox{d}w\,(w^2-1)^{3/2}\,
    \frac{2|V_j(w)|^2}{\Big(\beta_j^2-\frac{w+1}{2}\Big)^5}\,
    < \frac{3072\pi^2}{n_f}\,m_{B^*} m_{D^*}\,
    \widetilde\chi_{1^-}(0) \,, \nonumber\\[0.6cm]
   \sum_{j=1,2,3}&&
    \int\limits_{-\infty}^{-1}\!\mbox{d}w\,(w^2-1)^{3/2}\,
    (1+\delta_{j3})\,
    \frac{\beta_j^2|P_j(w)|^2}{\Big(\beta_j^2-\frac{w+1}{2}\Big)^4}
    < \frac{256\pi^2}{n_f}\,\widetilde\chi_{0^-}(0) \,,
    \nonumber\\[0.6cm]
   \sum_{j=1,2,3,4}&&
    \int\limits_{-\infty}^{-1}\!\mbox{d}w\,(w^2-1)^{1/2}
    \left( \frac{w+1}{2} \right)^2
    \frac{2|A_j(w)|^2}{\Big(\beta_j^2-\frac{w+1}{2}\Big)^4}
    \nonumber\\
   \mbox{}+ \sum_{j=5,6,7}&&
    \int\limits_{-\infty}^{-1}\!\mbox{d}w\,(w^2-1)^{1/2}
    \left( \frac{w+1}{2} \right)^2\,(1+\delta_{j7})\,
    \frac{(\beta_j^2-1)|A_j(w)|^2}
     {\Big(\beta_j^2-\frac{w+1}{2}\Big)^5} \nonumber\\
    &&\mbox{}< \frac{768\pi^2}{n_f}\,m_{B^*} m_{D^*}\,\chi_{1^+}(0) \,,
\label{newineq}
\end{eqnarray}
where we have set $Q^2=0$ for simplicity. In principle, the
inequalities could be strengthened by using negative values of $Q^2$,
which however must be sufficiently far away from threshold in order for
the operator product expansion to be well defined. However, in
Ref.~\cite{CaNe} it was found that this would improve the bounds only
slightly. Therefore, we shall not consider this possibility any
further.

In terms of the kinematical variable $w$ entering the unitarity
inequalities, the momentum transfer $t=q^2$ appearing in (\ref{pimunu})
and (\ref{chidr}) is given by
\begin{equation}
   t = q^2 = m_{B^{(*)}}^2 + m_{D^{(*)}}^2
   - 2 m_{B^{(*)}} m_{D^{(*)}} w \,.
\label{eq:momtransf}
\end{equation}
We work with $w$ instead of $t$ because it is the natural variable for
implementing heavy-quark symmetry, and because the various $B^{(*)}
D^{(*)}$ thresholds, which occur at different values of $t$, all occur
at the same value $w=-1$, enabling a unified treatment of all form
factors. The form factors $V_i$, $S_i$, $A_i$ and $P_i$, which appear
in (\ref{newineq}), are linear combinations of the traditonal
heavy-quark basis form factors chosen (i) to reduce to the Isgur--Wise
function in the heavy-quark limit, (ii) to have definite spin--parity
quantum numbers ($1^-$, $0^+$, $1^+$ and $0^-$, respectively), and
(iii) to reduce the unitarity relations to sums of squares. The
definitions of these form factors are given in the Appendix. The factor
of $n_f$ appears in (\ref{newineq}) because SU$(n_f)$ light-flavour
multiplets of heavy-meson states contribute with the same weight to the
unitarity sums. To be conservative, we take $n_f=2.5$ to account for
the breaking of SU(3) flavour symmetry due to the strange-quark mass.
For each $\bar B^{(*)}\to D^{(*)}$ transition form factor, we define a
quantity
\begin{equation}
   \beta_j = \frac{m_{B^{(*)}}+m_{D^{(*)}}}
              {2\sqrt{m_{B^{(*)}} m_{D^{(*)}}}} \,,
\label{beta}
\end{equation}
where it is understood that the appropriate masses are used. Moreover,
we use the notation
\begin{eqnarray}
   \widetilde\chi_{1^-}(0) &=& \chi_{1^-}(0)
    - \sum_{n=1,2} \frac{f_n^2(B_c^*)}{M_n^4(B_c^*)} \,,
    \nonumber\\
   \widetilde\chi_{0^-}(0) &=& \chi_{0^-}(0)
    - \sum_{n=1,2} \frac{f_n^2(B_c)}{M_n^2(B_c)} \,,
\end{eqnarray}
and thus include the contributions of the first two vector $B^*_c$ and
scalar $B_c$ mesons to the polarization functions.

For the vector correlator, the leading terms in the operator product
expansion of the vacuum polarization functions are given by
\begin{eqnarray}
   \chi_{0^+}(Q^2) &=& \frac{3}{4\pi^2} \int\limits_0^1\!
    \mbox{d}x\,x(1-x)\,\frac{x m_b^2 + (1-x) m_c^2 - m_b m_c}
    {x m_b^2 + (1-x) m_c^2 + x(1-x) Q^2} \,, \nonumber\\
   \chi_{1^-}(Q^2) &=& \frac{3}{8\pi^2} \int\limits_0^1\!
     \mbox{d}x\,x^2(1-x)^2\,
    \frac{3x m_b^2 + 3(1-x) m_c^2 + m_b m_c + 2x(1-x) Q^2}
    {\big[ x m_b^2 + (1-x) m_c^2 + x(1-x) Q^2 \big]^2} \,,
\label{chiqcd}
\end{eqnarray}
where $m_b$ and $m_c$ are the heavy-quark masses. The functions
$\chi_{0^-}$ and $\chi_{1^+}$ for the axial correlator are obtained
from these expressions by the replacement $m_c\to -m_c$. Evaluating
these results at $Q^2=0$, and using $z=m_c/m_b=0.29\pm 0.03$ for the
ratio of the heavy-quark pole masses, we obtain: $\chi_{0^+}(0) =
4.40_{+0.58}^{-0.52}\times 10^{-3}$, $m_b^2\,\chi_{1^-}(0) =
9.92_{+0.17}^{-0.19}\times 10^{-3}$, $\chi_{0^-}(0) =
2.09_{-0.06}^{+0.05}\times 10^{-2}$, and $m_b^2\,\chi_{1^+}(0) =
6.06_{+0.28}^{-0.27}\times 10^{-3}$. For the spin-1 channels, we use
the value $m_b=(4.8\pm 0.2)\,\mbox{GeV}$. The $O(\alpha_s)$
corrections, which have been calculated in
Refs.~\cite{Broa}--\cite{Djou}, enhance these results by 29\%, 31\%,
14\%, and 29\%, respectively, where we use $\alpha_s=\alpha_s(m_b)=
0.22$. The leading nonperturbative corrections, proportional to the
quark and gluon condensates, are very small and can be neglected
\cite{Gene,Rein}. With these parameters, we obtain
\begin{eqnarray}
   \chi_{0^+}(0) &=& (5.69\pm 0.71)\times 10^{-3} \,,\nonumber\\
   m_{B^*} m_{D^*}\,\widetilde\chi_{1^-}(0)
   &=& (3.73\pm 0.51)\times 10^{-3} \,, \nonumber\\
   \widetilde\chi_{0^-}(0) &=& (1.46\pm 0.06)\times 10^{-2} \,,
    \nonumber\\
   m_{B^*} m_{D^*}\,\chi_{1^+}(0) &=& (3.63\pm 0.34)\times 10^{-3} \,.
\label{chivals}
\end{eqnarray}
In the $1^-$ and $0^-$ channels, we have subtracted the pole
contributions from the two lowest $B_c$ and $B_c^*$ mesons using the
values of the decay constants shown in Table~\ref{tab:masses}. In the
evaluation of the inequalities (\ref{newineq}), we choose to relate all
form factors to the reference form factor $V_1(w)$, which governs the
$\bar B\to D\,\ell\,\bar\nu$ decay rate in (\ref{BDrate}), and whose
normalization at zero recoil is $V_1(1)=0.98\pm 0.07$ \cite{LNN}.
Then, what enters the inequalities (\ref{newineq}) are the products of
the correlation functions divided by $n_f\,V_1^2(1)$. The unitarity
bounds become weaker the larger the values of the correlation functions
and the smaller the value of $V_1(1)$. Therefore, we simultaneously
lower the value of the form factor and increase the central values in
(\ref{chivals}) by one standard deviation. This leads to the
conservative upper bounds
\begin{eqnarray}
   \frac{\chi_{0^+}(0)}{n_f\,V_1^2(1)}
   &<& 3.1\times 10^{-3} \,,\qquad
    m_{B^*} m_{D^*}\,\frac{\widetilde\chi_{1^-}(0)}{n_f\,V_1^2(1)}
    < 2.0\times 10^{-3} \,, \nonumber\\
   \frac{\widetilde\chi_{0^-}(0)}{n_f\,V_1^2(1)}
   &<& 7.3\times 10^{-3} \,,\qquad
    m_{B^*} m_{D^*}\,\frac{\chi_{1^+}(0)}{n_f\,V_1^2(1)}
    < 1.9\times 10^{-3} \,,
\label{chis}
\end{eqnarray}
which we use in our analysis.

\section{Model-independent bounds on form factors}
\label{sec:3}

The four inequalities in (\ref{newineq}) can be used to derive bounds
on the corresponding form factors in the semileptonic domain. The
problem is brought into a canonical form by performing a conformal
mapping $w\to z(w,a)$, which transforms the cut $w$ plane onto the
interior of the unit disc $|z|<1$, such that the integrals in
(\ref{newineq}) become integrals around the unit circle
$z=e^{i\theta}$. This is achieved by defining
\begin{equation}
    z(w,a) = \frac{\sqrt{w+1}-\sqrt 2\,a}{\sqrt{w+1}+\sqrt 2\,a}
   \,;\quad a>0 \,,
\label{mapp}
\end{equation}
which maps the branch point $w=-1$ onto $z=-1$. The choice of the real
parameter $a$ will be discussed later.

As explained in previous works \cite{Bour}--\cite{CaMa}, one further
introduces a set of ``outer'' functions $\phi_j(z)$, i.e.\ functions
without zeros nor singularities inside the unit disc, such that the
unitarity inequalities take the form
\begin{equation}
   \frac{1}{2\pi} \int\limits_0^{2\pi}\!\mbox{d}\theta\,
   \sum_j\,\l|\phi_j(e^{i\theta})\,F_j(w)\r|^2 \leq 1 \,,
\end{equation}
where it is implied that $w$ is expressed in terms of $z=e^{i\theta}$
by means of (\ref{mapp}). Here $F_j$ is one of the scalar,
pseudoscalar, vector or axial-vector form factors, and $j$ runs over
the set of form factors in a given spin--parity channel. Along the unit
circle, the modulus squared of the outer functions are equal to the
weights in front of the corresponding form factors in (\ref{newineq}),
multiplied by the modulus of the Jacobian of the conformal
transformation (\ref{mapp}). The calculation of these functions is
straightforward. The weight functions appearing in the inequalities are
of the generic form
\begin{equation}
   N\,(w^2-1)^{3/2-n} \left( \frac{w+1}{2} \right)^{2n}
   \left( \beta^2 - \frac{w+1}{2} \right)^{-m} \,,
\end{equation}
where $N$ is a constant. The corresponding outer functions are given by
\begin{eqnarray}
   \phi(z) &=& \sqrt{2\pi N}\,\frac{4^{2-n} a^4}{(\beta+a)^m}\,
    \left( \frac{1+a}{2a} \right)^{3/2-n}\,(1+z)^{n+2}\,
    (1-z)^{m-9/2} \nonumber\\
   &&\mbox{}\times \left( 1-z\,\frac{\beta-a}{\beta+a} \right)^{-m}\,
    \left( 1-z\,\frac{1-a}{1+a} \right)^{3/2-n} \,.
\end{eqnarray}

\bigskip
In deriving bounds on the values or derivatives of the form factors
inside the unit disc, one must account for the effects of singularities
located below the physical thresholds. For simplicity, we discuss the
singularity structure in terms of the momentum transfer $q^2$ given in
(\ref{eq:momtransf}). For the form factors of interest, the thresholds
occur at the values of $\sqrt{q^2}$ at which the relevant $B^{(*)}
D^{(*)}$ pairs can begin to be produced, i.e.\ at $(m_B+m_D)\approx
7.15\,\mbox{GeV}$, $(m_B+m_{D^*})\approx 7.29\,\mbox{GeV}$,
$(m_{B^*}+m_D)\approx 7.19\,\mbox{GeV}$, or $(m_{B^*}+m_{D^*})\approx
7.33\,\mbox{GeV}$. Subthreshold singularities arise due to the coupling
to states with masses below these thresholds. On the one hand, there
are cuts produced by multi-particle states. (Anomalous thresholds are
seen not to appear by inspecting the relevant triangle diagrams at both
quark and hadron level.) For instance, the form factor $S_3$, which
appears in the description of $\bar B^*\to D^*$ transitions, has a
branch point due to its coupling to $B D$ pairs as well as to
two-particle states consisting of a $B_c$ meson and light hadrons
\cite{Boyd1}. However, these cuts are rather short. Their effects are
suppressed by phase space and, in the case of $B_c$ states, by the
Zweig rule. Once a model for the discontinuities across the cuts is
adopted, these subthreshold singularities can be included in the
dispersive bounds \cite{CaNe,Capr3}, and their effect is found to be
very small. We thus neglect these contributions in the sequel.

On the other hand, the form factors also have singularities due to
single-particle states. Since both ground-state and orbitally excited
$B_c$ states are expected to be very narrow \cite{Eich}, we shall
assume that they give poles located on the real axis. Our choice of
form factors is such that the scalar functions $S_j$ receive only
contributions from scalar particles, the vector functions $V_j$ only
from vector particles, etc., situated below the corresponding physical
thresholds. The masses of these bounds states have already been given
in Table~\ref{tab:masses}. The residues of the corresponding poles,
which are proportional to the couplings of $B_c$ states to $B^{(*)}
D^{(*)}$ pairs, are poorly known. Fortunately, dispersive bounds can
still be derived even in the presence of poles with unknown residues
\cite{Taro2,Capr1}. The optimal technique is to multiply each form
factor $F_j$ by a specific function $B_j(z)$ (a so-called Blaschke or
``inner'' function) with zeros at the positions of the poles, and with
unit modulus along the physical cut. The explicit form of the inner
functions is
\begin{equation}
   B_j(z) = \prod_n \frac{z-z_n^{(j)}}{1-z z_n^{(j)}} \,,
\end{equation}
where the index $n$ runs over all $B_c$ bound-states that couple to a
given form factor, and
\begin{equation}
   z_n^{(j)} =
   \frac{\sqrt{(m_{B^{(*)}}+m_{D^{(*)}})^2 - M_n^2}
         - 2a \sqrt{m_{B^{(*)}} m_{D^{(*)}}}}
        {\sqrt{(m_{B^{(*)}}+m_{D^{(*)}})^2 - M_n^2}
         + 2a \sqrt{m_{B^{(*)}} m_{D^{(*)}}}}
\label{zjp}
\end{equation}
is the image of the pole of mass $M_n$ in the $z$ plane. In terms of
these functions, the unitarity inequality for a given spin--parity
$J^P$ is written as
\begin{equation}
   \frac{1}{2\pi} \int\limits_0^{2\pi}\!\mbox{d}\theta\,
   \sum_j\,|f_j(e^{i\theta})|^2 \leq 1 \,,
\label{compact2}
\end{equation}
where
\begin{equation}
   f_j(z) = B_j(z)\,\phi_j(z)\,F_j(w(z))
\label{fj}
\end{equation}
defines analytic functions inside the unit disc. We can thus Taylor
expand the functions $f_j(z)$ in (\ref{compact2}) and find that
\begin{equation}
   \sum_j \sum_{n=0}^\infty\,\l|\frac{1}{n!}\,
   f_j^{(n)}(0) \r|^2 \le 1 \,,
\label{master}
\end{equation}
where $f_j^{(n)}(0)$ denotes the $n$-th derivative of $f_j(z)$ with
respect to $z$, evaluated at $z=0$. There is one such inequality for
each spin--parity channel. The derivatives with respect to $z$ in
(\ref{master}) can be calculated using (\ref{fj}), where $B_j(z)$ and
$\phi_j(z)$ are known functions. The derivatives of the form factors at
$z=0$ are related to the derivatives with respect to $w$ at the
kinematical point $w=w_0$, defined such that $z(w_0,a)=0$. For the form
factor $V_1(w)$, for instance, we have
\begin{eqnarray}
   \partial_z V_1(w(z))\Big|_{z=0} &=& -8 a^2\rho_1^2\,V_1(w_0) \,,
    \nonumber\\
   \partial_z^2 V_1(w(z))\Big|_{z=0} &=& (128 a^4 c_1
    - 32 a^2\rho_1^2)\,V_1(w_0) \,, \nonumber\\
   \partial_z^3 V_1(w(z))\Big|_{z=0} &=& (3072 a^6 d_1 + 1536 a^4 c_1
    - 144 a^2\rho_1^2)\,V_1(w_0) \,,
\label{connect}
\end{eqnarray}
where
\begin{equation}
   w(z) = 2 a^2 \left( \frac{1+z}{1-z} \right)^2 - 1 \,,
\label{wzrel}
\end{equation}
and the parameters $\rho_1^2$, $c_1$ and $d_1$ are defined by the
expansion
\begin{equation}
   V_1(w) = V_1(w_0)\,\Big[ 1 - \rho_1^2 (w-w_0) + c_1 (w-w_0)^2
   + d_1 (w-w_0)^3 + \dots \Big] \,.
\label{Fjexp}
\end{equation}
Of course, the values of these parameters will depend on the choice of
$w_0$, i.e.\ on the choice of the parameter $a$ in (\ref{mapp}).

In the heavy-quark limit, all functions $F_j(w)$ become identical in
the semileptonic region and equal to the Isgur--Wise form factor. In
order to incorporate corrections to that limit, we choose the vector
function $V_1(w)$ as a reference form factor and express the expansion
parameters of all other form factors in terms of the reference
parameters $\rho_1^2$, $c_1$ and $d_1$ defined above. Our choice of the
reference form factor is motivated by the fact that it is the physical
form factor describing the semileptonic decay $\bar B\to
D\,\ell\,\bar\nu$. The leading symmetry-breaking corrections to the
ratios of heavy-meson form factors have been analysed in great detail
(for a review, see e.g.\ Ref.~\cite{Review}). We write
\begin{equation}
   R_j(w) \equiv \frac{F_j(w)}{V_1(w)}
   \equiv A_j\,\Big[ 1 + B_j (w-w_0) + C_j (w-w_0)^2
   + D_j (w-w_0)^3 + \dots \Big] \,.
\label{Rjdef}
\end{equation}
The results for the parameters $A_j,\dots,D_j$ obtained by including
the leading short-distance and $1/m_Q$ corrections, for two different
choices of $a$ relevant to our discussion, are given in the Appendix.
Using these results, the derivatives of the form factors $F_j(w)$ can
be expressed in terms of those of the reference form factor $V_1(w)$.

\section{Zero-recoil expansion}
\label{sec:4}

A convenient choice of the parameter $a$ in (\ref{mapp}) is to set
$a=1$, in which case the conformal transformation $w\to z$ maps the
zero-recoil point $w=1$ onto the origin $z=0$ of the unit disc (i.e.\
$w_0=1$). With this choice, it is straightforward to obtain bounds on
the form factors and their derivatives at zero recoil. We first give
the allowed values for the slope $\rho_1^2$ and the curvature $c_1$,
and then consider the coefficient $d_1$ of the third-order term in
(\ref{Fjexp}). In order to see the effect of the corrections to the
heavy-quark limit, we present the results both in the case of exact
spin symmetry\footnote{In fact, we only equate the first three
derivatives of the various form factors, but give all masses, poles and
thresholds their physical values.}
and with the leading corrections taken into account. The effect of
uncertainties in the corrections to the heavy-quark limit will be
considered below.

\begin{figure}
\centerline{\epsfxsize=12cm\epsffile{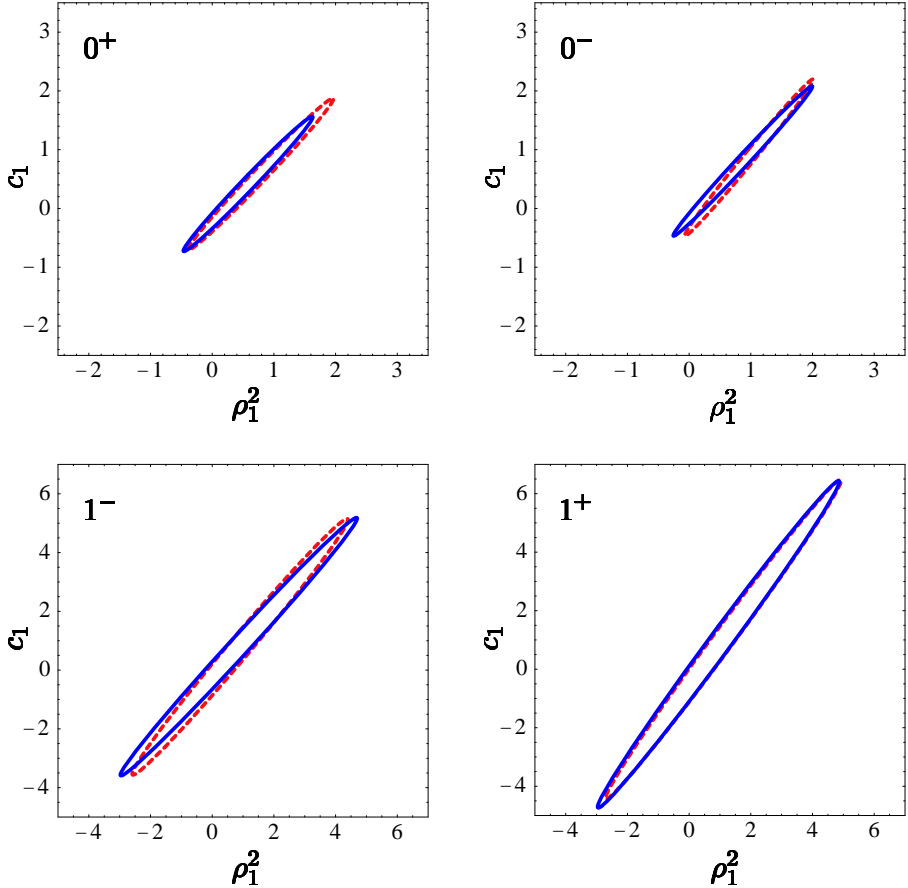}}
\centerline{\parbox{14cm}{\caption{\label{fig:ellipses}\small\sl
Bounds on the slope and curvature of the reference form factor $V_1(w)$
obtained from the unitarity inequalities in the different spin--parity
channels. Dashed curves correspond to the heavy-quark limit, while
solid ones include the leading corrections to that limit. Note that the
axes in the lower plots are extended by a factor of 2.}}}
\end{figure}

When only terms with up to two derivatives are kept in inequality
(\ref{master}), the resulting constraint on the parameters $\rho_1^2$
and $c_1$ of the reference form factor is given by an ellipse
\cite{CaNe}:
\begin{equation}
   (\rho_1^2 - \bar\rho_1^2)^2 + S\,\Big[ (c_1 - \bar c_1)
   - T\,(\rho_1^2 - \bar\rho_1^2) \Big]^2 < K^2 \,.
\label{ellipse}
\end{equation}
The single parameter sensitive to the values of the QCD correlators
given in (\ref{chis}) is the ``radius'' $K$. All other parameters of
the ellipse depend on known meson masses and, to a lesser extent, on
the symmetry-breaking corrections of (\ref{Rjdef}). In
Figure~\ref{fig:ellipses}, we show the resulting ellipses for the
different spin--parity channels, both with and without corrections to
the heavy-quark limit. Only values of the parameters inside the
ellipses are permitted. The effects of the corrections to the
heavy-quark limit are small and tend to align the major axes of the
four ellipses. In Table~\ref{tab:ellipses}, we give the corresponding
ellipse parameters, including the angle $\varphi$ of the major axis. In
all four cases, we find that $S\approx 40\mbox{--}60$, implying that
the ellipses are highly degenerate. As a consequence, the curvature
parameter $c_1$ is strongly correlated with the slope parameter
$\rho_1^2$, a model-independent property observed in Ref.~\cite{CaNe}.
It is apparent from Figure~\ref{fig:ellipses} that the bounds derived
from the $0^\pm$ channels are much stronger than (and fully contained
in) those derived from the $1^\pm$ channels. Therefore, we will not
consider the latter hereafter. For completeness, we note that the
(axial) vector and the (pseudo) scalar form factors are not fully
independent functions, since some of them have to obey kinematical
constraints at the point of maximum recoil, corresponding to $q^2=0$.
When deriving the dispersive bounds in the different spin--parity
channels, the kinematical constraints can be incorporated using
Lagrange multipliers. The dispersive bounds derived in the (pseudo)
scalar channel are then interconnected with those derived in the
(axial) vector channel, yielding a domain for the parameters of the
reference form factor that is smaller than the one obtained from the
intersection of the allowed domains obtained sperately from the
different spin--parity channels. In practice, however, the (pseudo)
scalar channel yields so much more restrictive bounds than the (axial)
vector channel that the resulting improvement is negligible.

\begin{table}
\centerline{\parbox{14cm}{\caption{\label{tab:ellipses}\small\sl
Parameters of the ellipses for the different spin--parity channels,
including corrections to the heavy-quark limit}}}
\vspace{0.4cm}
\begin{center}
\begin{tabular}{c|cccccc}
\hline\hline
$J^P$ & $\bar\rho_1^2$ & $\bar c_1$ & $T$ & $S$ & $K$ & $\varphi$\\
\hline
$0^+$ & 0.58 & 0.42 & 1.08 & 44.3 & 1.05 & $47.5^\circ$ \\
$0^-$ & 0.87 & 0.81 & 1.12 & 61.9 & 1.13 & $48.4^\circ$ \\
$1^-$ & 0.86 & 0.80 & 1.14 & 62.9 & 3.84 & $48.9^\circ$ \\
$1^+$ & 0.95 & 0.85 & 1.42 & 39.0 & 3.92 & $55.0^\circ$ \\
\hline\hline
\end{tabular}
\end{center}
\end{table}

To see what happens when the term with three derivatives is included,
consider the simpler case of only a single form factor contributing to
the unitarity sum. Then the corresponding inequality can be written in
a form similar to (\ref{ellipse}):
\begin{equation}
   (\rho_1^2 - b_1)^2 + (8 c_1 - b_2 - b_3\rho_1^2)^2
   + (64 d_1 - b_4 - b_5\rho_1^2 - b_6 c_1)^2 < K^2 \,,
\label{ellipsoid}
\end{equation}
where the $b_i$ are numerical coefficients. This relation describes a
highly degenerate ellipsoid in $(\rho_1^2,c_1,d_1)$ space, whose
axes are roughly in the ratios $1:1/8:1/64$. The coefficient $d_1$ is
thus determined within a rather small range:
\begin{equation}
   d_1 = \frac{b_4 + b_5\rho_1^2 + b_6 c_1}{64} \pm \Delta \,,
\label{d0sol}
\end{equation}
where $\Delta < K/64\approx 0.02$. We will see below that this is much
smaller than the range due to the theoretical uncertainty in the
relations between the various form factors. What changes in the general
case of many form factors are the precise values of the coefficients 8
and 64 in front of $c_1$ and $d_1$, as well as the values of the other
ellipsoid parameters. In addition, the inclusion of the third
derivative in (\ref{master}) reduces the allowed domain for the first
two derivatives. This feature is specific to the simultaneous treatment
of several form factors whose derivatives are related, as they are here
through heavy-quark symmetry. However, the inclusion of many form
factors does not change the strong degeneracy of the ellipsoid in
$(\rho_1^2,c_1,d_1)$ space.

\begin{figure}
\centerline{\epsfxsize=12cm\epsffile{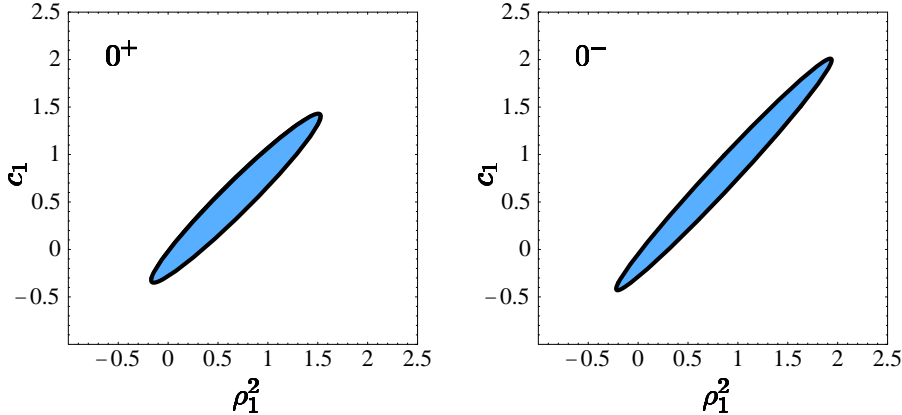}}
\centerline{\parbox{14cm}{\caption{\label{fig:ell3rd}\small\sl
Bounds on the slope and curvature of the reference form factor $V_1(w)$
obtained by including the third-derivative terms in the unitarity
inequalities for the $0^\pm$ channels, taking theoretical uncertainties
in the calculation of corrections that break heavy-quark symmetry into
account.
}}}
\end{figure}

\begin{table}
\centerline{\parbox{14cm}{\caption{\label{tab:finalell}\small\sl
Parameters of the ellipses obtained from the third-order inequalities,
including theoretical uncertainties. We also quote results for the
third-order coefficient $d_1=\alpha\rho_1^2+\beta c_1+\gamma$.}}}
\vspace{0.4cm}
\begin{center}
\begin{tabular}{c|cccccc|ccc}
\hline\hline
$J^P$ & $\bar\rho_1^2$ & $\bar c_1$ & $T$ & $S$ & $K$ & $\varphi$ &
 $\alpha$ & $\beta$ & $\gamma$\\
\hline
$0^+$ & 0.68 & 0.54 & 1.01 & 14.9 & 0.85 & $46.2^\circ$ &
 0.47(11) & $-1.33(6)$ & 0.01(11) \\
$0^-$ & 0.86 & 0.79 & 1.11 & 29.2 & 1.08 & $48.4^\circ$ &
 0.44(10) & $-1.37(6)$ & 0.05(11) \\
Combined & 0.67 & 0.55 & 1.03 & 25.9 & 0.84 & $46.4^\circ$ &
 0.45(7)\phantom{1} & $-1.35(4)$ & 0.03(8)\phantom{1} \\
\hline\hline
\end{tabular}
\end{center}
\end{table}

At first sight, the inclusion of the third derivative terms bring an
improvement which appears to be very significant. For the case of the
$0^+$ channel, for instance, the resulting allowed region for the
parameters $\rho_1^2$ and $c_1$ is reduced, with respect to the one
shown in Figure~\ref{fig:ellipses}, by more than a factor of 3.
However, the situation changes if we allow for theoretical errors
$\delta B_j$, $\delta C_j$ and $\delta D_j$ in the coefficients
entering the relations (\ref{Rjdef}) between the various form
factors.\footnote{There is no need to include errors in the
normalization of the form fatcors at zero recoil, since we have already
used a conservative low value for the normalization of the reference
form factor.}
To this end, we generate a large number of ellipses by taking random
values of the theoretical errors within intervals $[-\delta B,\delta
B]$, $[-\delta C,\delta C]$, and $[-\delta D,\delta D]$, setting
$\delta B=\delta C=\delta D=0.1$, which we believe is a conservative
estimate of the uncertainties. The results are shown in
Figure~\ref{fig:ell3rd}. Whereas the effect of theoretical
uncertainties is moderate in the case of the second-order inequality,
it has a large impact at third order. The envelope of the ellipses
obtained in that case is much larger than each single ellipse, and
almost fills up the ellipse obtained at second order. This shows that
not much can be gained by going to yet higher orders in the expansion.
The thick ellipses in Figure~\ref{fig:ell3rd} are a good approximation
to the allowed regions. The parameters of these ellipses are given in
Table~\ref{tab:finalell}. Their intersection is again to excellent
approximation given by an ellipse, whose parameters are given in the
last row. We also present the results for the third-order expansion
coefficient $d_1$, which we write as $d_1=\alpha\rho_1^2 + \beta c_1 +
\gamma$. We find that the allowed range for this parameter obtained
from (\ref{d0sol}) is only little affected by theoretical errors. The
consistency between the results obtained from the two channels allows
us to take the average of the two results, shown in the last row.

\begin{figure}
\centerline{\epsfxsize=10.75cm\epsffile{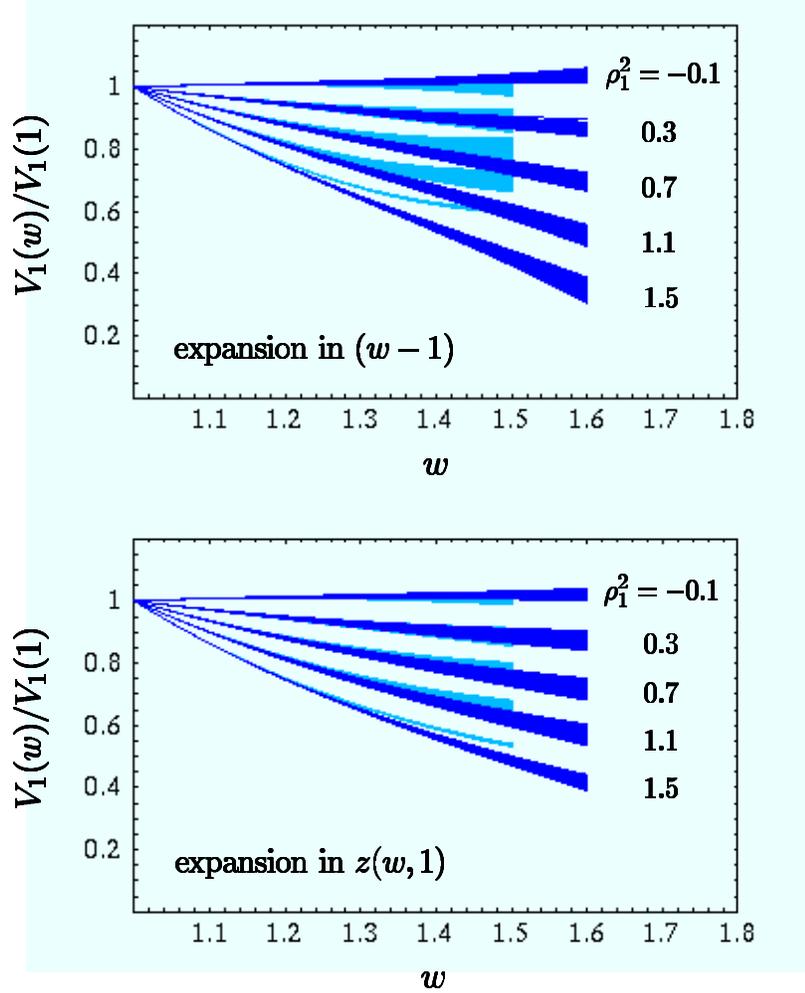}}
\centerline{\parbox{14cm}{\caption{\label{fig:Gfuns}\small\sl
Allowed shapes of the function $V_1(w)$ for various values of
$\rho_1^2$. The dark (light) bands correspond to the third-order
(second-order) expansion of the form factor. In the upper plot, the
expansion is performed in powers of $(w-1)$, in the lower plot in
powers of the variable $z(w,1)$.
}}}
\end{figure}

To exhibit the possible behaviour of the form factor $V_1(w)$ predicted
by our results, we show in the upper plot in Figure \ref{fig:Gfuns} the
allowed shapes for this function for a selection of equally spaced
values of the slope parameter $\rho_1^2$ inside the allowed region,
given by $-0.17<\rho_1^2<1.51$. For each choice of $\rho_1^2$, the
parameters $c_1$ and $d_1$ are scanned over the regions allowed by the
dispersive bounds. The results are shown with and without including the
third-order term in the zero-recoil expansion, i.e.\ the expansion in
(\ref{Fjexp}) with $w_0=1$. Close to zero recoil (i.e.\ for $1\le
w<1.2$), both approximations provide an equally accurate description of
the form factor. For larger recoil, the effect of the third-order term
becomes clearly visible. The convergence rate of the series can be
improved by expressing the form factor $V_1(w(z))$, with $w(z)$ given
in (\ref{wzrel}), as a power series in $z$ rather than in $(w-1)$. The
corresponding results are shown in the lower plot in
Figure~\ref{fig:Gfuns}. Now the differences between the second- and
third-order expansions become important only at larger values of $w$,
and we are confident that the third-order results provide an adequate
representation of the form factor over the enitre kinematical region
accessible in semileptonic decays. Because the spread in the curves
caused by variations of the parameter $c_1$ and $d_1$ inside the
allowed regions is very small, our results can be represented, to a
very good approximation, by taking the values of these parameters along
the major axes. This gives the one-parameter function
\begin{eqnarray}
   \frac{V_1(w)}{V_1(1)} &\approx& 1 - 8 \rho_1^2 z
    + (51.\rho_1^2 - 10.) z^2 - (252.\rho_1^2-84.) z^3 \,, \nonumber\\
   \mbox{with}\quad
   &z& = \frac{\sqrt{w+1}-\sqrt 2}{\sqrt{w+1}+\sqrt 2} \,.
\label{V13rd}
\end{eqnarray}
As an aside, we note that expanding this result in powers of $(w-1)$
would yield the slope--curvature relation $c_1\approx
1.05\rho_1^2-0.15$, which is not very different from the relation
$c_1\approx 0.74\rho_1^2-0.09$ obtained in Ref.~\cite{CaNe} with some
assumptions about subthreshold singularities. We stress, however, that
in the present work such assumptions are avoided.

\section{Optimized expansion}
\label{sec:5}

We have seen above that higher-order terms in the expansion of the form
factor $V_1(w)$ around the zero-recoil point have a significant effect
already for rather small values of $w$. The convergence of the
expansion was improved considerably by introducing the variable $z$
instead of $(w-1)$; however, the question arises whether an even better
convergence can be achieved by optimizing the choice of the expansion
variable.

When one wishes to approximate a function along an interval, the most
natural expansion is given in terms of orthonormal polynomials on this
interval. The domain of convergence of the expansion is then an ellipse
passing through the first singularity of the function \cite{Walsh}. In
order to accelerate the rate of convergence in the interval of
interest, one must conformally map the interior of the function's
analyticity domain onto the inside of an ellipse, such that the
interval of interest is applied to the segment between the focal points
\cite{CuCi}. This conformal mapping is given by elliptic functions
\cite{Nehari}. However, when the physical interval is far from the
first singularity (as it is for semileptonic $\bar B^{(*)}\to D^{(*)}$
decays), the ellipse is almost degenerate to a circle, so that a simple
Taylor expansion about the center of the ellipse will have almost the
same radius of convergence as the expansion in orthonormal polynomials
discussed above. In the absence of subthreshold singularities, this
approximate circle is simply a rescaling of the circle obtained using
the conformal mapping of (\ref{mapp}), with $a$ determined such that
the physical region is applied onto an interval symmetric about the
origin $z=0$. This gives $a=a_{\rm opt}$ such that $z(w_{\rm
max},a_{\rm opt})=-z(1,a_{\rm opt})$, which is very close to the
variable first suggested by Boyd and Lebed \cite{BoLeb}. For the case
of $\bar B\to D$ transitions, we find $a_{\rm opt}\approx 1.067$ and
$z_{\rm max}\equiv z(w_{\rm max},a_{\rm opt})\approx 0.032$.

The argument just presented is only valid when there are no
subthreshold singularities, while the form factors of interest have
poles and branch points below the physical cut. Nevertheless, because
the singularities are far from the semileptonic domain, the variable
$z(w,a_{\rm opt})$ remains an excellent choice. Indeed, a measure of
the rate of convergence of an expansion of a form factor in a variable
$x$ is provided by the ratio $r_{\rm conv}=|x|_{\rm max}/|x_{\rm
pole}|$ where $x_{\rm pole}$ is the position of the first pole and
$|x|_{\rm max}$ the largest value $|x|$ can take in the semileptonic
domain. The form factor $S_1(w)$, for instance, has a pole due to a
scalar state at $6.70\,\mbox{GeV}$, so that $r_{\rm conv}\approx 0.07$.
For comparison, we note that $r_{\rm conv}\approx 0.35$ for the
expansion variable $x=(w-1)$, and $r_{\rm conv}\approx 0.15$ for the
variable $x=z(w,1)$ used in the previous section. It is clear that
$z(w,a_{\rm opt})$ gives the best rate of convergence. In order to
simplify notation in what follows, we will write $z_*$ for $z(w,a_{\rm
opt})$ and $a_*$ for $a_{\rm opt}$. We shall use the same value of
$a_*$ for all form factors.

Our claim is that all $\bar B^{(*)}\to D^{(*)}$ form factors can be
accurately described by a second-order polynomial in $z_*$. In
particular, we shall derive dispersive constraints on the slope
$\rho_{1*}^2$ and the curvature $c_{1*}$ (now taken at the recoil point
$w=w_0\approx 1.276$) in the expansion of the reference form factor
$V_1(w(z_*))$, i.e.
\begin{equation}
   \frac{V_1(w)}{V_1(w_0)} \approx  1 - 8 a_*^2 \rho_{1*}^2 z_*
    + 16 a_*^2 (4 a_*^2 c_{1*} - \rho_{1*}^2) z_*^2 \,.
\label{eq:ffparam}
\end{equation}
To estimate the truncation error due to the neglect of higher-order
terms, consider the expansion of the analytic functions $f_j(z)$ of
(\ref{fj}) around $z=0$. If we write
\begin{equation}
   f_j(z) = f_j(0) \left( 1 + \frac{f_j^{(1)}(0)}{f_j(0)}\,z
   + \frac{f_j^{(2)}(0)}{f_j(0)}\,\frac{z^2}{2} + \Delta_j(z)
   \right) \,,
\end{equation}
the remainder $\Delta_j(z)$ can be bounded using the dispersive
constraint (\ref{master}). We find that
\begin{equation}
   |\Delta_j(z)| < |z|^3 \sqrt{\frac{1}{|f_j(0)|^2} - 1} \,.
\end{equation}
Let us concentrate, for concreteness, on the scalar form factor
$S_1(w)$. Making the conservative assumption that $S_1(w_0)$ is as low
as $\frac 23\,S_1(1)$, we find $|\Delta_{S_1}(z)| < 7\times 10^{-4}$
for all values of $z$ corresponding to the semileptonic region. This
truncation error is tiny and will be neglected it in what follows. The
second step consists in noticing that the known function
$[\phi_{S_1}(z_*) B_{S_1}(z_*)]^{-1}$ is well approximated by a
second-order polynomial in $z_*$. This is where the convergence
criterion discussed earlier comes into play, since $1/B_{S_1}(z_*)$
contains the scalar poles. The fact that $r_{\rm conv}=|z_*|_{\rm
max}/|z_{*{\rm pole}}|$ is so small guarantees that the expansion of
$[\phi_{S_1}(z_*) B_{S_1}(z_*)]^{-1}$ converges rapidly. We find that
the first three terms in this expansion deviate by at most 0.2\% from
the full function over the whole semileptonic region. As a result, the
form factor $S_1(w)$, which is proportional to the product of the two
quantities $f_{S_1}(z_*)$ and $[\phi_{S_1}(z_*) B_{S_1}(z_*)]^{-1}$, is
approximated to an accuracy of better than about 0.2\% by the product
of the truncated expansions of these two factors. Of course, this
product includes terms of order $z_*^3$ and $z_*^4$. However, their
contribution never exceeds 1.5\% over the full semileptonic region (it
is lower than 1\% in the region $w<1.5$). Thus, we conclude that the
scalar form factor $S_1(w)$ can be described by a second-order
polynomial in $z_*$ with an accuracy of better than 2\%. A similar
conclusion holds for all other $\bar B\to D^{(*)}$ form factors, since
they are related to the $S_1(w)$ by heavy-quark symmetry, up to small
corrections.

As in Section~\ref{sec:4}, we use the dispersive bounds to constrain
the first two derivatives of $V_1(w)$ at $w_0$. However, $w_0$ is not
the zero-recoil point here, and the left-hand side of (\ref{master})
contains as an overall factor the unknown normalization $|V_1(w_0)|^2$.
To proceed, we multiply both sides of the inequality by a common
factor,
\begin{equation}
   \left| \frac{V_1(1)}{V_1(w_0)} \right|^2 \sum_j
   \sum_{n=0}^\infty\,\left| \frac{1}{n!}\,f_j^{(n)}(0) \right|^2
   \le \left| \frac{V_1(1)}{V_1(w_0)} \right|^2 \,,
\label{master2}
\end{equation}
such that now the left-hand side contains the known normalization
$V_1(1)$. To evaluate the right-hand side of (\ref{master2}) we rely on
our observation that in the semileptonic region all $\bar B^{(*)}\to
D^{(*)}$ form factors are very well described by a second-order
polynomial in $z_*$, and thus use (\ref{eq:ffparam}) with $w=1$. This
approximation enables us to write the resulting constraints on
$\rho_{1*}^2$ and $c_{1*}$ as ellipses, like in the zero-recoil case.

\begin{figure}
\centerline{\epsfxsize=12cm\epsffile{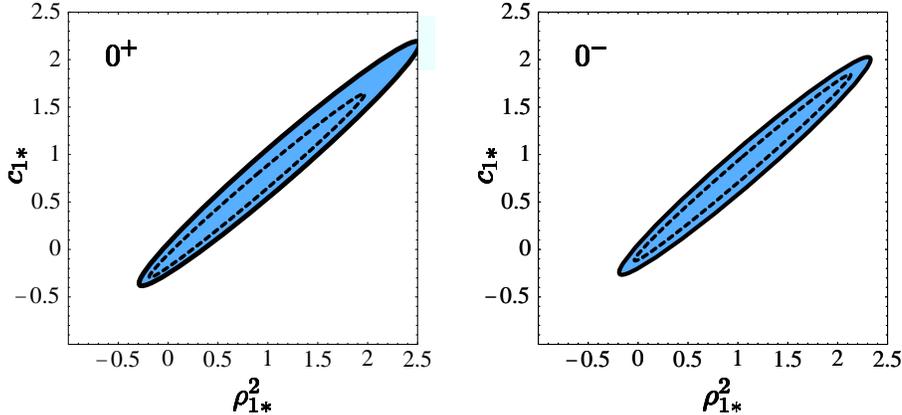}}
\centerline{\parbox{14cm}{\caption{\label{fig:ellopt}\small\sl
Bounds on the slope and curvature of the reference form factor $V_1(w)$
obtained using an expansion in the optimized variable $z_*$. The solid
(dashed) ellipses show the results obtained with (without) including
theoretical uncertainties in the heavy-quark relations.
}}}
\end{figure}

\begin{table}
\centerline{\parbox{14cm}{\caption{\label{tab:ell_opt}\small\sl
Parameters of the ellipses obtained from the second-order inequalities,
using the optimized variable $z_*$
}}}
\vspace{0.4cm}
\begin{center}
\begin{tabular}{c|cccccc}
\hline\hline
$J^P$ & $\bar\rho_1^2$ & $\bar c_1$ & $T$ & $S$ & $K$ & $\varphi$ \\
\hline
$0^+$ & 1.11 & 0.89 & 0.90 & 29.0 & 1.40 & $42.5^\circ$ \\
$0^-$ & 1.07 & 0.88 & 0.89 & 27.6 & 1.26 & $42.2^\circ$ \\
\hline\hline
\end{tabular}
\end{center}
\end{table}

For the scalar and pseudoscalar channels, the ellipses obtained in the
$(\rho_{1*}^2,c_{1*})$ plane are shown in Figure~\ref{fig:ellopt}. The
dashed ones correspond to using the central values of the parameters
$A_j$, $B_j$ and $C_j$ defined in (\ref{Rjdef}) and given in the
Appendix. As in Section~\ref{sec:4}, we account for theoretical
uncertainties in the values of these parameters by generating a large
number of ellipses, allowing $B_j$ and $C_j$ to vary by $\pm 0.1$. This
procedure generates the shaded areas, which to a good approximation are
contained in the larger ellipses. We find again a similarity between
the final ellipses in the two channels and, in particular, a near
alignement of their major axes. The intersection of these two ellipses
is, to a very good approximation, given by the pseudoscalar ellipse
alone. This is in contrast to the zero-recoil expansion considered in
the previous section, where it was the scalar channel that gave the
tighter constraints. For completeness, the ellipse parameters are given
in Table~\ref{tab:ell_opt}.

\begin{figure}
\centerline{\epsfxsize=10cm\epsffile{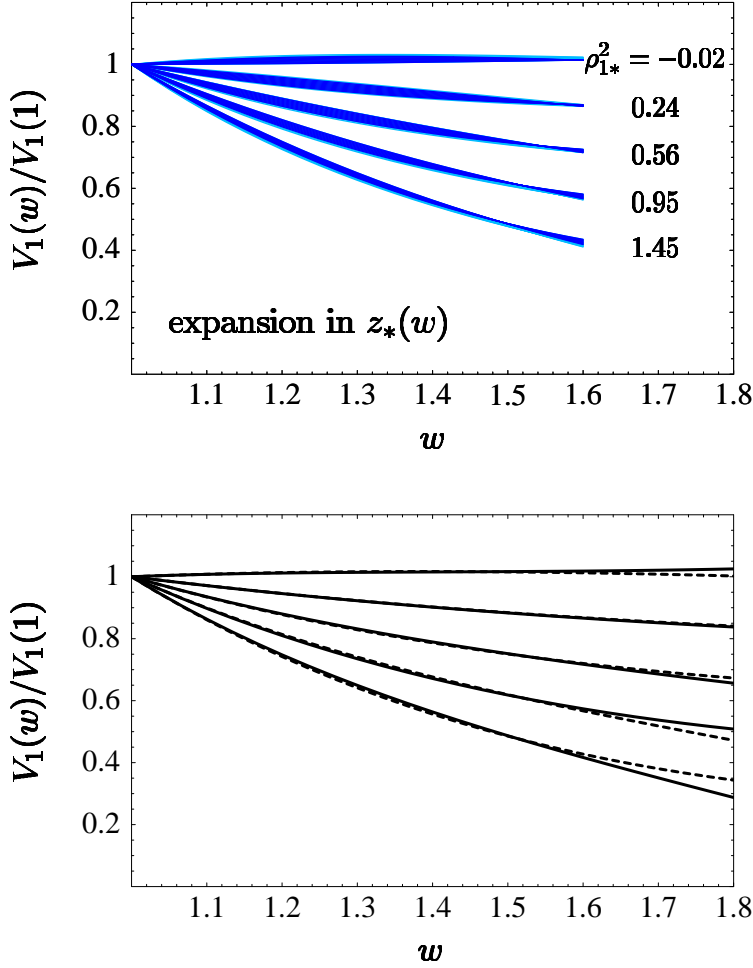}}
\centerline{\parbox{14cm}{\caption{\label{fig:Gfuns_opt}\small\sl
Upper plot: Allowed shapes for the function $V_1(w)$ for various values
of $\rho_{1*}^2$ (the slope at $w_0\approx 1.276$) chosen such that
they correspond to the values of $\rho_1^2$ (the slope at $w=1$) of
Figure~\protect\ref{fig:Gfuns}. The light bands include an estimate of
the truncation error. Lower plot: Comparison of the two one-parameter
approximations (\protect\ref{V13rd}) (solid) and (\protect\ref{V1opt})
(dashed).
}}}
\end{figure}

Let us now investigate the properties of the optimized parametrization
and compare them with the results obtained in the previous section. In
the upper plot in Figure~\ref{fig:Gfuns_opt}, we show the allowed
shapes of the form factor $V_1(w)$ for values of the slope parameter
$\rho_{1*}^2$ chosen such that they correspond to the values of the
zero-recoil slope $\rho_1^2$ used in Figure~\ref{fig:Gfuns}. For each
choice of $\rho_{1*}^2$, the parameter $c_{1*}$ is scanned over the
region allowed by the dispersive bounds. Because $S$ is large compared
with $K^2$, and because the term multiplying $c_{1*}$ in expression
(\ref{eq:ffparam}) for the form factor is suppressed by the small
factor $z_*^2$, the spread in the curves caused by the variation of
$c_{1*}$ is very small. Therefore, to a very good accuracy, $c_{1*}$
can be replaced by $\bar c_1 + \tan\varphi\,(\rho_{1*}^2 -
\bar\rho^2_1)$. This gives the one-parameter function
\begin{eqnarray}
   \frac{V_1(w)}{V_1(1)} &\approx&
   \frac{1 - 9.105\rho_{1*}^2 z_* + (57.0\rho_{1*}^2-7.5) z_*^2}
        {0.354\rho_{1*}^2 + 0.992} \,, \nonumber\\
   \mbox{with}\quad
   &z_*& = \frac{\sqrt{w+1}-1.509}{\sqrt{w+1}+1.509} \,.
\label{V1opt}
\end{eqnarray}
The light-shaded bands in the figure include, in addition to the spread
discussed above, an estimate of the truncation error due to the neglect
of higher-order terms in $z_*$. It is obtained by not expanding the
known functions $[\phi_j(z_*) B_j(z_*)]^{-1}$ in powers of $z_*$, but
keeping their full expressions to calculate the form factors from the
second-order expansions of the functions $f_j(z_*)$. (We have argued
above that the error from the truncation of the functions $f_j(z_*)$
themselves is a negligible effect.) These bands reflect the total
theoretical uncertainty in our results, which is seen never to exceed
the level of 2\%. As a consequence, the one-parameter description of
the form factor given above can be trusted to that level of precision.

In the lower plot in Figure~\ref{fig:Gfuns_opt}, we compare the
parametrization (\ref{V1opt}) to the one given in (\ref{V13rd}). The
parameters $\rho_1^2$ and $\rho_{1*}^2$ are chosen as before. We find
that the two sets of curves are nearly indistinguishable in the
semileptonic region $1\le w<1.6$, whereas they start to diverge for
larger $w$ values. This agreement provides a strong consistency check
of our method, given that the expansions in the two variables $z$ and
$z_*$ are carried out to different order, and that the dispersive
bounds in the two cases are dominated by different spin--parity
channels. We thus conclude that, for all practical purposes, both
parametrizations are equally reliable. Similar parametrizations for all
other $\bar B^{(*)}\to D^{(*)}$ form factors can be obtained using
(\ref{Rjdef}) and the results given in the Appendix.

\section{Phenomenological applications}
\label{sec:more}

We now summarize our results and provide simple, constrained
parametrizations for the form factors which determine the differential
rates for $\bar B\to D\,\ell\,\bar\nu$ and $\bar B\to
D^{*}\ell\,\bar\nu$ decays. The form factor $V_1(w)$ governing the
first process is described to an accuracy of better than 2\% by the two
parametrizations given in (\ref{V13rd}) and (\ref{V1opt}). The two
slope parameters entering these expressions are related by
\begin{equation}
   \rho_{1*}^2 \approx \frac{0.62\rho_1^2+0.04}{1-0.22\rho_1^2}
    \,,\qquad
   \rho_1^2 \approx
   \frac{1.61\rho_{1*}^2-0.06}{1+0.36\rho_{1*}^2} \,.
\end{equation}
The dispersive bounds discussed in Section~\ref{sec:4} require that
$-0.17<\rho_1^2<1.51$, and therefore $-0.07<\rho_{1*}^2<1.47$. These
intervals, by themselves, represent nontrivial bounds on the slope of
the form factor, which are competitive with the most recent bounds
derived from inclusive heavy-quark sum rules \cite{Ira97}.

To obtain similar parametrizations of the function ${\cal F}(w)$, which
governs the $\bar B\to D^*\ell\,\bar\nu$ decay rate in (\ref{BDrate}),
we find it convenient to first relate this function to the axial-vector
form factor $A_1(w)$ through \cite{Review}
\begin{eqnarray}
   \left[ 1 + \frac{4w}{w+1}\,\frac{1-2w r+r^2}{(1-r)^2} \right]\,
   {\cal F}(w)^2 &=& \Bigg\{ 2\,\frac{1-2w r+r^2}{(1-r)^2} \left[
    1 + \frac{w-1}{w+1}\,R_1(w)^2 \right] \nonumber\\
   &&\mbox{}+ \left[ 1 + \frac{w-1}{1-r}\Big( 1 - R_2(w)\Big)
    \right]^2 \Bigg\}\,A_1(w)^2 \,,
\end{eqnarray}
where $r=m_{D^*}/m_B$, and $R_1(w)$ and $R_2(w)$ are ratios of form
factors defined in the Appendix, and are given by
\begin{eqnarray}
   R_1(w) &=& \frac{h_V(w)}{h_{A_1}(w)}
    \approx 1.27 - 0.12(w-1) + 0.05(w-1)^2 \,, \nonumber\\
   R_2(w) &=& \frac{h_{A_3}(w)+r h_{A_2}(w)}{h_{A_1}(w)}
    \approx 0.80 + 0.11(w-1) - 0.06(w-1)^2 \,.
\end{eqnarray}
The theoretical predictions for these ratios are supported by
measurements reported by the CLEO Collaboration \cite{CLEOff}. We have
chosen to work with $A_1(w)$ instead of ${\cal F}(w)$, because the
latter suffers from large kinematical enhancements of the corrections
to the heavy-quark limit. In other words, the coefficients in the
expansion of the ratio ${\cal F}(w)/V_1(w)$ corresponding to
(\ref{A1V1}) below would be large. These large corrections were not
accounted for in Ref.~\protect\cite{Boyd3}, where the symmetry-breaking
corrections in the relation between ${\cal F}(w)$ and $A_1(w)$ are
neglected.

Using the results of the Appendix, the form factor $A_1(w)$ can be
related to the reference form factor $V_1(w)$ through
\begin{eqnarray}
   \frac{A_1(w)}{V_1(w)} &\approx& 0.948 \Big( 1 - 0.212 z
    - 4.007 z^2 - 1.342 z^3 + \dots \Big) \nonumber\\
   &\approx& 0.937 \Big( 1 - 0.476 z_* - 4.163 z_*^2 + \dots \Big)\,,
\label{A1V1}
\end{eqnarray}
where $z=z(w,1)$ and $z_*=z(w,a_*)$. From this, it is straightforward
to derive the parametrizations of $A_1(w)$ from the results for
$V_1(w)$ given in (\ref{V13rd}) and (\ref{V1opt}). They are
\begin{eqnarray}
   \frac{A_1(w)}{A_1(1)} &\approx& 1 - 8\rho_{A_1}^2 z
    + (53.\rho_{A_1}^2 - 15.) z^2 - (231.\rho_{A_1}^2-91.) z^3 \,,
   \nonumber\\
   \mbox{with}\quad
   &z& = \frac{\sqrt{w+1}-\sqrt 2}{\sqrt{w+1}+\sqrt 2} \,,
\label{A13rd}
\end{eqnarray}
and
\begin{eqnarray}
   \frac{A_1(w)}{A_1(1)} &\approx&
   \frac{1 - 9.105\rho_{A_1 *}^2 z_* + (61.3\rho_{A_1 *}^2-14.8) z_*^2}
        {0.358\rho_{A_1 *}^2 + 0.984} \,, \nonumber\\
   \mbox{with}\quad
   &z_*& = \frac{\sqrt{w+1}-1.509}{\sqrt{w+1}+1.509} \,.
\label{A1opt}
\end{eqnarray}
The relation between the two slope parameters entering these
expressions is
\begin{equation}
   \rho_{A_1 *}^2 \approx
   \frac{0.60\rho_{A_1}^2+0.07}{1-0.22\rho_{A_1}^2} \,,\qquad
   \rho_{A_1}^2 \approx
   \frac{1.66\rho_{A_1 *}^2-0.12}{1+0.36\rho_{A_1 *}^2} \,.
\end{equation}
The dispersive bounds discussed in Section~\ref{sec:4} require that
$-0.14<\rho_{A_1}^2<1.54$ and $-0.01<\rho_{A_1 *}^2<1.51$.
Alternatively, we could repeat the derivation of the dispersive bounds
using $A_1(w)$ instead of $V_1(w)$ as the reference form factor. Since
the symmetry-breaking corrections in the relation between the two form
factors are very small, the two procedures give essentially the same
results.

\begin{figure}
\centerline{\epsfxsize=10cm\epsffile{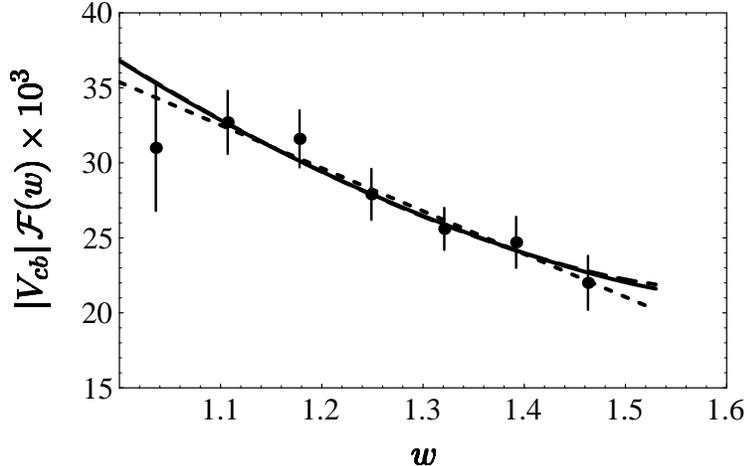}}
\centerline{\parbox{14cm}{\caption{\label{fig:vcb}\small\sl
Experimental data for the product $|V_{cb}|\,{\cal F}(w)$ as a function
of $w$, extracted from semileptonic $\bar B\to D^*\ell\,\bar\nu$ decays
\protect\cite{cleovcb}. The solid curve is a fit using our
parametrization in (\protect\ref{A13rd}), while the (barely visible)
long-dashed one refers to the parametrization of (\protect\ref{A1opt}).
The two curves are nearly indistinguishable in the semileptonic domain.
The short-dashed line shows a straight-line fit, similar to the one
used in the experimental analysis.
}}}
\end{figure}

\begin{table}
\centerline{\parbox{14cm}{\caption{\label{tab:vcbfit}\small\sl
Results of the various fits to experimental data on the recoil spectra
in $\bar B\to D^*\ell\,\bar\nu$ decays \protect\cite{cleovcb}. The
slope parameters are measured at $w=1$.
}}}
\vspace{0.4cm}
\begin{center}
\begin{tabular}{l|ccc}
\hline\hline
Parametrization & $|V_{cb}|\,{\cal F}(1)$ & $\rho_{A_1}^2$ &
 $\rho_{\cal F}^2$ \\
\hline
Eq.~(\protect\ref{A13rd}) & 0.037(2) & 1.36(18) & 1.15(18) \\
Eq.~(\protect\ref{A1opt}) & 0.037(2) & 1.34(17) & 1.13(17) \\
Linear Fit & 0.035(2) & --- & 0.81(13) \\
\hline\hline
\end{tabular}
\end{center}
\end{table}

Our results will eliminate the large uncertainty associated with the
extrapolation of the experimental recoil spectra in $\bar B\to
D^{(*)}\ell\,\bar\nu$ decays to zero recoil. At present, this
uncertainty is the main experimental error in the determination of
$|V_{cb}|$ \cite{Gibbons}. In fact, in many experimental analyses a
linear shape of the form factor has been used to extrapolate the data,
without attributing an error to this assumption. A reanalysis using our
constrained parametrizations is therefore highly desirable in order to
get a reliable determination of $|V_{cb}|$. To illustrate the potential
impact of our results, we show in Figure~\ref{fig:vcb} different fits
to the data for the product ${\cal F}(w)\,|V_{cb}|$ measured in $\bar
B\to D^*\ell\,\bar\nu$ decays by the CLEO Collaboration \cite{cleovcb}.
Similar measurements have also been reported in
Refs.~\cite{ARGUS}--\cite{OPAL}. Our goal here is not to establish a
new value of $|V_{cb}|$ (which is best done by the experimental
collaborations themselves), but rather to illustrate how our
parametrizations work. The short-dashed line shows a linear fit to the
data, whereas the solid and long-dashed curves show the fits obtained
using our parametrizations. All fits have an excellent
$\chi^2/\mbox{dof}$ between 0.3 and 0.4, reflecting that we are
neglecting corrrelations between the errors of the various data points.
Clearly, the two parametrizations we propose give nearly
indistinguishable fits. More importantly, the obtained values of
$|V_{cb}|$ and the zero-recoil slope parameters are systematically
larger than those derived from the linear fits.

The results of the fits are summarized in Table~\ref{tab:vcbfit}, where
we also quote values for the slopes of the form factors $A_1(w)$ and
${\cal F}(w)$ at zero recoil. The theoretical prediction for the
difference between the two slope parameters, which can be derived using
the results of the Appendix, is $\rho_{\cal
F}^2\approx\rho_{A_1}^2-0.21$ \cite{Review}.

\section{Conclusions}
\label{sec:ccl}

We have derived dispersive bounds on the shape of the form factors
describing semileptonic $\bar B\to D^{(*)} \ell\,\bar\nu$ decays. The
method we use is based on dispersion relations and complex-analysis
techniques, which combine QCD calculations of current--current
correlation functions in the Euclidean domain with spectral
representations of these functions in terms of sums over intermediate
hadronic states. Using the positivity of these spectral functions and
crossing symmetry, it is possible to derive bounds on the form factors
of particular hadron states that are contained in the spectral sum. In
the case of heavy mesons, these bounds can be strengthened by including
in the unitarity sum all contributions that are related, in the
semileptonic region, by heavy-quark spin symmetry. Thus, we include the
contributions of $B D$, $B D^*$, $B^* D$ and $B^* D^*$ states, taking
into account the leading short-distance and $1/m_Q$ corrections to the
heavy-quark limit. We consider all four spin--parity channels relevant
for $\bar B^{(*)}\to D^{(*)}$ transitions and combine them in an
optimal way.

We first apply these methods to derive bounds on the slope, curvature
and third derivative of a reference form factor in an expansion around
the zero-recoil point $w=1$, using as input the value of this form
factor at zero recoil. We find that a consistent inclusion of the
third-derivative term requires a careful treatment of theoretical
uncertainties in the relations that connect various form factors in the
heavy-quark limit. We show explicitely how accounting for these
uncertainties reduces significantly the apparent improvement brought by
the third-derivative term.

In analysing the different spin--parity channels, we find that the
scalar and pseudoscalar channels give significantly better bounds than
the vector and axial-vector ones. We find a strong correlation between
the curvature and slope of the reference form factor, and an even
stronger correlation of the third derivative with the slope and
curvature. This observation has led us to investigate expansions in
other kinematical variables, in which the strong correlations between
the parameters of the $(w-1)$ expansion are explained naturally. We
find that the variable $z(w,a)$ defined in (\ref{mapp}) accomplishes
this task, and furthermore leads to an improved convergence of the
expansion of the form factors. We have discussed two choices of the
parameter $a$ that are particularly convenient: the choice $a=1$ leads
to an expansion around zero recoil, however in a better variable than
$(w-1)$; the choice $a=a_*$ discussed in Section~\ref{sec:5}
corresponds to an expansion around a kinematical point close to the
center of the semileptonic region, thus providing nearly optimal
convergence. For the first choice, we find that the scalar channel
provides the tightest constraints, and that the form factors can be
well approximated by third-order polynomials. For the second choice,
the pseudoscalar channel gives the best bounds, and the form factors
can be approximated by second-order polynomials. In both cases, we
derive simple one-parameter functions for the physical form factors
$V_1(w)$ and $A_1(w)$, which are accurate to better than 2\% in the
semileptonic region. Similar parametrizations for all other $\bar
B^{(*)}\to D^{(*)}$ form factors can be obtained using (\ref{Rjdef})
and the results given in the Appendix. The fact that the
parametrizations obtained in the two cases lead to shapes of the form
factors that are nearly indistinguishable in the semileptonic region
provides a strong consistency check of the method. We note that our
functions are simpler than those derived by Boyd et al.\ \cite{Boyd3}
and include the leading short-distance and $1/m_Q$ corrections to the
heavy-quark limit, as well as the uncertainties in their calculation.
Nevertheless, we have checked that in fits to available experimental
data our results for the form factors agree with those obtained in
Ref.~\cite{Boyd3} to better than 3\% for values $w<1.5$. For larger
values of $w$, the parametrization for ${\cal F}(w)$ given in this
reference becomes less reliable.

For the convenience of the reader, we restate our results obtained in
the zero-recoil expansion based on the variable $z(w,1)$, which gives
the simplest parametrizations. They are
\begin{eqnarray}
   \frac{V_1(w)}{V_1(1)} &\approx& 1 - 8 \rho_1^2 z
    + (51.\rho_1^2 - 10.) z^2 - (252.\rho_1^2-84.) z^3 \,, \nonumber\\
   \frac{A_1(w)}{A_1(1)} &\approx& 1 - 8\rho_{A_1}^2 z
    + (53.\rho_{A_1}^2 - 15.) z^2 - (231.\rho_{A_1}^2-91.) z^3 \,,
\end{eqnarray}
where $z=(\sqrt{w+1}-\sqrt{2})/(\sqrt{w+1}+\sqrt{2})$, and $\rho_1^2$
and $\rho_{A_1}^2$ are the slope parameters at zero recoil, restricted
to the intervals $-0.17<\rho_1^2<1.51$ and $-0.14<\rho_{A_1}^2<1.54$.
The values of the form factors at zero recoil are known to be
$V_1(1)=0.98\pm 0.07$ and $A_1(1)=0.91\pm 0.03$. In
Section~\ref{sec:more}, we have shown how these results can be used to
determine $|V_{cb}|$ from an extrapolation of experimental data on
semileptonic decays to zero recoil. Not only do our parametrizations
eliminate the main experimental uncertainty in this extraction, but
they also lead to a corrected value for $|V_{cb}|$ that is
systematically higher (by about one standard deviation) than the value
obtained from a linear fit.

\bigskip
{\it Acknowledgments:\/}
One of us (I.C.) is grateful to the CERN Theory Division, and to Centre
de Physique des Particules de Marseille, for the hospitality extended
to her during part of this work. We thank the authors of
Ref.~\cite{Boyd3} for communicating results of their work prior to
publication, and David Rousseau for useful discussions.

\newpage
\setcounter{equation}{0}
\renewcommand{\theequation}{A.\arabic{equation}}
\setcounter{table}{0}
\renewcommand{\thetable}{A.\arabic{table}}
\section*{Appendix: Form factors and heavy-quark relations}

In the context of the heavy-quark expansion, it is common practice to
parametrize the meson matrix elements of the weak currents $V^\mu=\bar
c\gamma^\mu b$ and $A^\mu=\bar c\gamma^\mu\gamma^5 b$ by a set of form
factors $h_i(w)$ that depend on the kinematical variable $w=v\cdot v'$,
where $v$ and $v'$ are the meson velocities. With a mass-independent
normalization of meson states, these functions are defined as
\cite{FaNe}
\begin{eqnarray}
   \la D(v')|V^\mu|\bar B(v)\ra &=& h_+(w)\,(v+v')^\mu
    + h_-(w)\,(v-v')^\mu \,, \nonumber\\
   \la D^*(v',\eps')|V^\mu|\bar B(v)\ra
   &=& i h_V(w)\,\epsilon^{\mu\nu\alpha\beta}\,
    \eps_\nu^{\prime*} v'_\alpha v_\beta \,, \nonumber\\
   \la D(v')|V^\mu|\bar B^*(v,\eps)\ra
   &=& i h_{\bar V}(w)\,\epsilon^{\mu\nu\alpha\beta}\,
    \eps_\nu v'_\alpha v_\beta \,, \nonumber\\
   \la D^*(v',\eps')|V^\mu|\bar B^*(v,\eps)\ra
   &=& - \Big[ h_1(w)\,(v+v')^\mu + h_2(w)\,(v-v')^\mu \Big]\,
    \eps^{\prime*}\!\cdot\eps \nonumber\\
   &&\mbox{}+ h_3(w)\,\eps^{\prime*}\!\cdot v\,\eps^\mu
    + h_4(w)\,\eps\cdot v'\,\eps^{\prime*\mu} \nonumber\\
   &&\mbox{}- \Big[ h_5(w)\,v^\mu + h_6(w)\,v'^\mu \Big]\,
    \eps^{\prime*}\!\cdot v\,\eps\cdot v' \,,
\label{defv}
\end{eqnarray}
for the vector current, and
\begin{eqnarray}
   \la D^*(v',\eps')|A^\mu|\bar B(v)\ra &=& h_{A_1}(w)\,(w+1)\,
    \eps^{\prime*\mu} - \Big[ h_{A_2}(w)\,v^\mu + h_{A_3}(w)\,v'^\mu
    \Big]\,\eps^*\!\cdot v \,, \nonumber\\
   \la D(v')|A^\mu|\bar B^*(v,\eps)\ra &=& h_{\bar A_1}(w)\,(w+1)\,
    \eps^\mu - \Big[ h_{\bar A_2}(w)\,v'^\mu + h_{\bar A_3}(w)\,v^\mu
    \Big]\,\eps\cdot v' \,, \nonumber\\
   \la D^*(v',\eps')|A^\mu|\bar B^*(v,\eps)\ra
   &=& i\epsilon^{\mu\nu\alpha\beta} \bigg\{
    \Big[ h_7(w)\,(v+v')^\mu + h_8(w)\,(v-v')^\mu \Big]\,
    \eps_\alpha \eps_\beta^{\prime *} \nonumber\\
   &&\mbox{}+ \Big[ h_9(w)\,\eps^{\prime *}\!\cdot v\,\eps_\nu
    + h_{10}(w)\,\eps\cdot v'\,\eps_\nu^{\prime *} \Big]\,
    v'_\alpha v_\beta \bigg\} \,,
\label{defa}
\end{eqnarray}
for the axial-vector current. Here, $\eps^{(\prime)}$ denote the
polarization of the vector mesons. For our purposes, it is convenient
to introduce a new set of form factors defined as linear combinations
of the functions $h_i(w)$. We introduce these form factors in such a
way that they reduce to the Isgur--Wise function in the heavy-quark
limit, have definite spin--parity quantum numbers, and ``diagonalize''
the the unitarity relations derived in Section~\ref{sec:2}. Thus, we
define the scalar ($J^P=0^+$) functions (dropping the argument $w$ for
brevity)
\begin{eqnarray}
   S_1^{(BD)} &=& h_+ - \frac{1+r}{1-r}\,\frac{w-1}{w+1}\,h_- \,,
    \nonumber\\
   S_2^{(B^* D^*)} &=& h_1 - \frac{1+r}{1-r}\,\frac{w-1}{w+1}\,h_2
    \,, \nonumber\\
   S_3^{(B^* D^*)} &=& w \left( h_1 - \frac{1+r}{1-r}\,
    \frac{w-1}{w+1}\,h_2 \right) \nonumber\\
   &&\mbox{}+ \frac{w-1}{1-r}\,\Big[ r\,h_3 - h_4 + (1-w r)\,h_5
    + (w-r)\,h_6 \Big] \,,
\end{eqnarray}
the pseudoscalar ($J^P=0^-$) functions
\begin{eqnarray}
   P_1^{(B D^*)} &=& \frac{1}{1+r}\,\Big[ (w+1)\,h_{A_1}
    - (1-w r)\,h_{A_2} - (w-r)\,h_{A_3} \Big] \,, \nonumber\\
   P_2^{(B^* D)} &=& \frac{1}{1+r}\,\Big[ r(w+1)\,h_{\bar A_1}
    - (r-w)\,h_{\bar A_2} - (r w-1)\,h_{\bar A_3} \Big] \,,
    \nonumber\\
   P_3^{(B^* D^*)} &=& h_7 - \frac{1-r}{1+r}\,h_8 \,,
\end{eqnarray}
the vector ($J^P=1^-$) functions
\begin{eqnarray}
   V_1^{(B D)} &=& h_+ - \frac{1-r}{1+r}\,h_- \,, \nonumber\\
   V_2^{(B^* D^*)} &=& h_1 - \frac{1-r}{1+r}\,h_2 \,, \nonumber\\
   V_3^{(B^* D^*)} &=& w \left( h_1 - \frac{1-r}{1+r}\,h_2 \right)
    \nonumber\\
   &&\mbox{}+ \frac{1}{1+r}\,\Big[ (1-w r)\,h_3 + (r-w)\,h_4
    + (w^2-1)\,(r h_5 + h_6) \Big] \,, \nonumber\\
   V_4^{(B D^*)} &=& h_V \,, \nonumber\\
   V_5^{(B^* D)} &=& h_{\bar V} \,, \nonumber\\
   V_6^{(B^* D^*)} &=& h_3 \,, \nonumber\\
   V_7^{(B^* D^*)} &=& h_4 \,,
\end{eqnarray}
and the axial-vector ($J^P=1^+$) functions
\begin{eqnarray}
   A_1^{(B D^*)} &=& h_{A_1} \,, \nonumber\\
   A_2^{(B^* D)} &=& h_{\bar A_1} \,, \nonumber\\
   A_3^{(B^* D^*)} &=& h_7 - \frac{w-1}{w+1}\,h_8 + (w-1)\,h_{10}
    \,, \nonumber\\
   A_4^{(B^* D^*)} &=& h_7 + \frac{w-1}{w+1}\,h_8 + (w-1)\,h_9 \,,
    \nonumber\\
   A_5^{(B D^*)} &=& \frac{1}{1-r}\,\Big[ (w-r)\,h_{A_1}
    - (w-1)\,(r h_{A_2} + h_{A_3}) \Big] \,, \nonumber\\
   A_6^{(B^* D)} &=& \frac{1}{1-r}\,\Big[ (1-w r)\,h_{\bar A_1}
    + (w-1)\,(h_{\bar A_2} + r h_{\bar A_3}) \Big] \,, \nonumber\\
   A_7^{(B^* D^*)} &=& h_7 - \frac{1+r}{1-r}\,\frac{w-1}{w+1}\,h_8 \,.
\end{eqnarray}
In these equations, $r=m_{D^{(*)}}/m_{B^{(*)}}$ denotes the appropriate
ratio of meson masses. The functions $h_{\bar V}$ and $h_{\bar A_i}$,
which have not been defined in Ref.~\cite{FaNe}, are obtained from the
corresponding functions $h_V$ and $h_{A_i}$ by an interchange of the
heavy-quark masses, $m_c\leftrightarrow m_b$.

Luke's theorem protects the functions $h_+$, $h_{A_1}$, $h_{\bar A_1}$,
$h_1$, and $h_7$ from receiving first-order power corrections at zero
recoil \cite{Luke}. Up to corrections of order $1/m_Q^2$, they obey the
normalization conditions
\begin{equation}
   S_j(1) = \eta_V \,, \qquad A_j(1) = \eta_A \,,
\end{equation}
where $\eta_V$ and $\eta_A$ are the QCD renormalization constants of
the vector and axial currents at zero recoil \cite{Review}. This
implies that, to order $1/m_Q$, the functions appearing in the
unitarity inequalities for the spin--parity channels $0^+$ and $1^+$ in
(\ref{newineq}) have the same normalization.

In the heavy-quark limit, $h_+ = h_V = h_{A_1} = h_{A_3} = h_{\bar V} =
h_{\bar A_1} = h_{\bar A_3} = h_1 = h_3 = h_4 = h_7 = \xi$, where
$\xi(w)$ is the universal Isgur--Wise function, while all other
functions $h_i$ vanish. The leading short-distance corrections to this
limit can be parametrized in terms of the Wilson coefficients $C_i$ and
$C_i^5$ of the vector and axial currents ($i=1,2,3$), which are
calculable functions of the variable $w$ and the heavy-quark masses,
$m_b$ and $m_c$. For our purposes, it is sufficient to work with the
exact one-loop expressions for these functions \cite{Pasc}--\cite{QCD}.
Leading logarithms, which have been summed to all orders in
perturbation theory, are universal and cancel in the form-factor ratios
we are interested in. In fact, for the same reason we are free to
divide out $C_1$ as a common factor. Then the resulting one-loop
expressions are
\begin{eqnarray}
   \frac{C_1^5}{C_1} &=& 1 - \frac{4\alpha_s}{3\pi}\,r(w) \,,
    \nonumber\\
   \frac{C_2^{(5)}}{C_1} &=& -\frac{2\alpha_s}{3\pi}\,
    H_{(5)}(w,1/z) \,, \nonumber\\
   \frac{C_3^{(5)}}{C_1} &=& \mp\frac{2\alpha_s}{3\pi}\,
    H_{(5)}(w,z) \,,
\end{eqnarray}
where $z=m_c/m_b$, and
\begin{eqnarray}
   r(w) &=& \frac{1}{\sqrt{w^2-1}}\,\ln\left( w + \sqrt{w^2-1}
    \right) \,, \nonumber\\
   H_{(5)}(w,z) &=& \frac{z(1 - \ln z \mp z)}{1-2wz+z^2}
    + \frac{z}{\big(1-2wz+z^2\big)^2}\,\bigg\{
    2(w\mp 1)z(1\pm z)\,\ln z \nonumber\\
   &&\mbox{}- \left[ (w\pm 1) - 2w(2w\pm 1)z + (5w\pm 2w^2\mp 1)z^2
    - 2z^3 \right]\,r(w) \bigg\} \,.
\end{eqnarray}
The upper signs refer to the function $H$, the lower ones to $H_5$.

Similarly, the leading power corrections can be parametrized by six
functions $L_i(w)$ introduced in Ref.~\cite{FaNe}, which are related to
the subleading Isgur--Wise functions $\eta(w)$ and $\chi_i(w)$ defined
in Ref.~\cite{Luke}. Once again, since we are only interested in
form-factor ratios, we can omit the contribution of $\chi_1(w)$, which
is the same for all meson form factors. Then then resulting expressions
are
\begin{eqnarray}
   L_1 &=& -4(w-1)\,\frac{\chi_2}{\xi} + 12\,\frac{\chi_3}{\xi}
    \approx 0.72(w-1)\bar\Lambda \,, \nonumber\\
   L_2 &=& -4\,\frac{\chi_3}{\xi}\approx -0.16(w-1)\bar\Lambda \,,
    \nonumber\\
   L_3 &=& 4\,\frac{\chi_2}{\xi}\approx -0.24\bar\Lambda\,,
    \nonumber\\
   L_4 &=& (2\eta-1)\,\bar\Lambda\approx 0.24\bar\Lambda \,,
    \nonumber\\[0.15cm]
   L_5 &=& -\bar\Lambda \,, \nonumber\\
   L_6 &=& -\frac{2}{w+1}\,(\eta+1)\,\bar\Lambda
    \approx -\frac{3.24}{w+1}\,\bar\Lambda \,,
\end{eqnarray}
where $\bar\Lambda=m_M-m_Q\approx 0.5\,\mbox{GeV}$ is the ``binding
energy'' of a heavy meson, and on the right-hand side we have used
approximate expressions for the subleading Isgur--Wise functions
obtained using QCD sum rules \cite{Yossi}.

The corresponding expressions for the functions $h_i$ are
\begin{eqnarray}
   \frac{h_+}{\xi} &=& C_1 + \frac{w+1}{2}\,(C_2 + C_3)
    + (\eps_c + \eps_b)\,L_1 \,, \nonumber\\
   \frac{h_-}{\xi} &=& \frac{w+1}{2}\,(C_2 - C_3)
    + (\eps_c - \eps_b)\,L_4 \,, \nonumber\\
   \frac{h_V}{\xi} &=& C_1
    + \eps_c\,(L_2 - L_5) + \eps_b\,(L_1 - L_4) \,, \nonumber\\
   \frac{h_{A_1}}{\xi} &=& C_1^5
    + \eps_c\,\left( L_2 - \frac{w-1}{w+1}\,L_5 \right)
    + \eps_b\,\left( L_1 - \frac{w-1}{w+1}\,L_4 \right) \,,
    \nonumber\\
   \frac{h_{A_2}}{\xi} &=& C_2^5
    + \eps_c\,(L_3 + L_6) \,, \nonumber\\
   \frac{h_{A_3}}{\xi} &=& C_1^5 + C_3^5
    + \eps_c\,(L_2 - L_3 - L_5 + L_6) + \eps_b\,(L_1 - L_4) \,,
    \nonumber\\
   \frac{h_{\bar V}}{\xi} &=& C_1
    + \eps_c\,(L_1 - L_4) + \eps_b\,(L_2 - L_5) \,, \nonumber\\
   \frac{h_{\bar A_1}}{\xi} &=& C_1^5
    + \eps_c\,\left( L_1 - \frac{w-1}{w+1}\,L_4 \right)
    + \eps_b\,\left( L_2 - \frac{w-1}{w+1}\,L_5 \right) \,,
    \nonumber\\
   \frac{h_{\bar A_2}}{\xi} &=& - C_3^5
    + \eps_b\,(L_3 + L_6) \,, \nonumber\\
   \frac{h_{\bar A_3}}{\xi} &=& C_1^5 - C_2^5
    + \eps_c\,(L_1 - L_4) + \eps_b\,(L_2 - L_3 - L_5 + L_6) \,,
    \nonumber\\
   \frac{h_1}{\xi} &=& C_1 + \frac{w+1}{2}\,(C_2 + C_3)
    + (\eps_c + \eps_b)\,L_2 \,, \nonumber\\
   \frac{h_2}{\xi} &=& \frac{w+1}{2}\,(C_2 - C_3)
    + (\eps_c - \eps_b)\,L_5 \,, \nonumber\\
   \frac{h_3}{\xi} &=& C_1
    + \eps_c\,\left[ L_2 + (w-1)\,L_3 + L_5 - (w+1)\,L_6 \right]
    + \eps_b\,(L_2 - L_5) \,, \nonumber\\
   \frac{h_4}{\xi} &=& C_1
    + \eps_c\,(L_2 - L_5)
    + \eps_b\,\left[ L_2 + (w-1)\,L_3 + L_5 - (w+1)\,L_6 \right]
    \,, \nonumber\\
   \frac{h_5}{\xi} &=& - C_2
    + \eps_c\,(L_3 - L_6) \,, \nonumber\\
   \frac{h_6}{\xi} &=& - C_3
    + \eps_b\,(L_3 - L_6) \,, \nonumber\\
   \frac{h_7}{\xi} &=& C_1^5 + \frac{w-1}{2}\,(C_2^5 - C_3^5)
    + (\eps_c + \eps_b)\,L_2 \,, \nonumber\\
   \frac{h_8}{\xi} &=& \frac{w+1}{2}\,(C_2^5 + C_3^5)
    + (\eps_c - \eps_b)\,L_5 \,, \nonumber\\
   \frac{h_9}{\xi} &=& - C_2^5
    + \eps_c\,(L_3 - L_6) \,, \nonumber\\
   \frac{h_{10}}{\xi} &=& C_3^5
    + \eps_b\,(L_3 - L_6) \,,
\end{eqnarray}
where $\eps_b=1/2 m_b$ and $\eps_c=1/2 m_c$. Given these explicit
expressions, it is straightforward to evaluate the form-factor ratios
defined in (\ref{Rjdef}) including the leading corrections to the
heavy-quark limit. In order to be consistent, we expand the expressions
for the ratios to linear order in $\alpha_s$ and $1/m_Q$. In
Table~\ref{tab:HQS}, we show the results for the expansion coefficients
in (\ref{Rjdef}), with $w_0=1$. In the numerical analysis, we use
$\alpha_s=\alpha_s(\sqrt{m_b m_c})=0.26$, $z=m_c/m_b=0.29$, and
$\bar\Lambda=0.48\,\mbox{GeV}$. In Table~\ref{tab:HQS2}, we give the
corresponding values for the same expansion parameters, but at
$w_0\approx 1.267$.

\begin{table}
\centerline{\parbox{14cm}{\caption{\label{tab:HQS}\small\sl
Theoretical predictions for the coefficients in the expansion of the
form-factor ratios $R_j(w)$ around $w_0=1$
}}}
\vspace{0.4cm}
\begin{center}
\begin{tabular}{l|rrrr}
\hline\hline
$F_j$ & $A_j~~$ & $B_j~~$ & $C_j~~$ & $D_j~~$ \\
\hline
$S_1$ & 1.0036 & $-0.0068$ &   0.0017  & $-0.0013$ \\
$S_2$ & 1.0036 & $-0.0355$ & $-0.0813$ &   0.0402 \\
$S_3$ & 1.0036 &   0.0776  & $-0.1644$ &   0.0817 \\
\hline
$P_1$ & 1.1548 & $-0.2088$ & 0.0032 & $-0.0009$ \\
$P_2$ & 0.9060 & $-0.0727$ & 0.0031 & $-0.0008$ \\
$P_3$ & 1.0325 & $-0.2116$ & 0.0032 & $-0.0009$ \\
\hline
$V_1$ & 1 & 0 & 0 & 0 \\
$V_2$ & 1.0681 & $-0.1944$ &   0.0000  & 0.0000 \\
$V_3$ & 1.1361 & $-0.2474$ &   0.0000  & 0.0000 \\
$V_4$ & 1.2179 & $-0.1428$ & $-0.0015$ & 0.0004 \\
$V_5$ & 1.0676 & $-0.0362$ & $-0.0015$ & 0.0004 \\
$V_6$ & 1.4919 & $-0.2278$ & $-0.0015$ & 0.0004 \\
$V_7$ & 1.3416 & $-0.1987$ & $-0.0015$ & 0.0004 \\
\hline
$A_1$ & 0.9484 & $-0.0265$ & $-0.0560$ &   0.0266 \\
$A_2$ & 0.9484 &   0.0050  & $-0.0184$ &   0.0078 \\
$A_3$ & 0.9484 & $-0.0205$ & $-0.0869$ &   0.0421 \\
$A_4$ & 0.9484 &   0.0256  & $-0.1245$ &   0.0609 \\
$A_5$ & 0.9484 &   0.2984  & $-0.2391$ &   0.1180 \\
$A_6$ & 0.9484 & $-0.1699$ &   0.0621  & $-0.0324$ \\
$A_7$ & 0.9484 & $-0.0113$ & $-0.0857$ &   0.0414 \\
\hline\hline
\end{tabular}
\end{center}
\end{table}

\begin{table}
\centerline{\parbox{14cm}{\caption{\label{tab:HQS2}\small\sl
Theoretical predictions for the coefficients in the expansion of some
of the form-factor ratios $R_j(w)$ around $w_0\approx 1.267$
}}}
\vspace{0.4cm}
\begin{center}
\begin{tabular}{l|rrr}
\hline\hline
$F_j$ & $A_j~~$ & $B_j~~$ & $C_j~~$ \\
\hline
$S_1$ & 1.0018 & $-0.0061$ & $0.0009$  \\
$S_2$ & 0.9883 & $-0.0727$ & $-0.0554$ \\
$S_3$ & 1.0140 &  $0.0025$ & $-0.1117$ \\
\hline
$P_1$ & 1.0974 & $-0.2072$ & $0.0026$ \\
$P_2$ & 0.8862 & $-0.0711$ & $0.0025$ \\
$P_3$ & 0.9743 & $-0.2100$ & $0.0025$ \\
\hline
$A_1$ & 0.9373 & $-0.0522$ & $-0.0387$ \\
\hline\hline
\end{tabular}
\end{center}
\end{table}

\end{document}